\documentclass[twocolumn]{aastex631}
\usepackage{natbib}
\usepackage[utf8]{inputenc}
\usepackage{soul}
\usepackage{amsmath}
\usepackage{txfonts}
\usepackage{graphics,graphicx,epsfig}
\usepackage[version=4]{mhchem}
\usepackage{xcolor}
\newcommand{\gcc}{g\,cm$^{-3}$}

\begin{document}

\title{Serpentinization in the thermal evolution of icy Kuiper belt objects in the early Solar system}

\author[0000-0001-5531-1381]{Anik\'o Farkas-Tak\'acs}
\affiliation{Konkoly Observatory, Research Centre for Astronomy and Earth Sciences, Konkoly Thege 15-17, H-1121~Budapest, Hungary} 
\affiliation{E\"otv\"os Lor\'and University, Faculty of Science, P\'azm\'any P. st. 1/A, 1171 Budapest, Hungary} 

\author[0000-0002-8722-6875]{Csaba~Kiss}
\affiliation{Konkoly Observatory, Research Centre for Astronomy and Earth Sciences, Konkoly Thege 15-17, H-1121~Budapest, Hungary} 
\affiliation{ELTE E\"otv\"os Lor\'and University, Institute of Physics, P\'azm\'any P. st. 1/A, 1171 Budapest, Hungary} 

\author[0000-0002-7039-8099]{S\'andor G\'obi}
\affiliation{E\"otv\"os Lor\'and University, Faculty of Science, P\'azm\'any P. st. 1/A, 1171 Budapest, Hungary} 
\affiliation{MTA-ELTE Lend\"ulet Laboratory Astrochemistry Research Group, Institute of Chemistry, ELTE E\"otv\"os Lor\'and University, H-1518 Budapest, Hungary} 

\author[0000-0001-6420-510X]{\'Akos Kereszturi}
\affiliation{Konkoly Observatory, Research Centre for Astronomy and Earth Sciences, Konkoly Thege 15-17, H-1121~Budapest, Hungary} 

\begin{abstract}
Here we present an improved algorithm to model the serpentinization process in planetesimals in the early Solar system. Although it is hypothesized that serpentinization-like reactions played an important role in the thermal evolution of planetesimals, few and restricted models are available in this topic. These process may be important as the materials involved were abundant in these objects. Our model is based on the model by \citet{Gobi2017}, and contains improvements in the consideration of heat capacities and lithospheric pressure, and in the calculation of the amount of interfacial water. Comparison of our results with previous calculations show that there are significant differences in the e.g. the serpentinization time -- the time necessary to consume most of the reactants at specific initial conditions -- or the amount of heat produced by this process. In a simple application we show that in icy bodies, under some realistic conditions, below the melting point of water ice, serpentinization reaction using interfacial water may be able to proceed and eventually push the local temperature above the melting point to start a 'runaway' serpentinization. According to our calculations in objects with radii R\,$\gtrsim$\,200\,km serpentinization might have quickly reformed nearly the whole interior of these bodies in the early Solar system. 
\end{abstract}
\keywords{planetary systems: planets and satellites: composition, interiors}

\section{Introduction}\label{intro}

We only have indirect information on the internal material properties of the small bodies in the Solar system, coming chiefly from the analysis of meteorites.
Some samples show that, in addition to heat from accretion and radioactive decay, there may have been chemical processes that at some point in the early history of the Solar system significantly altered the mineralogical and lithological characteristics of the objects and contributed to internal heat production.
One of these chemical processes is the hydration of silicates. As it is an exothermic reaction, it may have contributed to the heating of planetesimals in the early Solar system \citep{Gail2014}, especially in bodies with water ice content.
The aqueous alteration found in carbonaceous chondrites is partly explained by the formation of Mg–serpentine via the serpentinization and we use here serpentinization as a model approach, as it is a moderately simple reaction and uses abundant reactants. Serpentinite consists of one or more serpentine group minerals resulting from the hydration of silicates. The presence of serpentine has been detected in meteorites, especially in CM chondrites with high carbon content whose primary minerals are members of the serpentine group and account for 55–58\% of the meteorite by volume \citep{Scott1988}.

According to some views, serpentinization is a very rare process, and serpentine found inside meteorites were formed already in the planetary nebula or during accretion \citep{Lunine2006}. 
There may have also been other changes inside the planetesimals that could alter serpentine.
 For instance, members of the serpentine family may have been dehydrated and/or altered by heat.
Serpentinization itself can produce significant heat, which can cause serpentine to become amorphous, as it has already been found in some meteorites \citep{Zega2003}. This is probably the reason why no well-crystallized serpentine minerals have been observed.

There are several reactions to form serpentinite from olivine. In these reactions the rock absorbs a large amount of water and consequently destroys the structure of the original minerals while it increases its volume and decreases its density. 
In one of the main serpentinization processes Mg-pyroxenes (enstatite, \ce{MgSiO3}) is also required in addition to the Mg\,-\,rich olivine (forsterite, \ce{Mg2SiO4}); the end product of the reaction is purely serpentinite (\ce{Mg3Si2O5(OH)4}).
The following reaction shows the stoichiometric equation of the formation of serpentinite:
\begin{equation}
\ce{Mg2SiO4 + MgSiO3 + 2H2O -> Mg3Si2O5(OH)4}
\label{eq:pyrox}
\end{equation}
\noindent This reaction takes place only in the presence of liquid water at a temperature- and pressure-dependent reaction rate \citep{Wegner1983}.
In this case, the enthalpy is $69\,kJ\,mol^{-1}$ \citep{Robie1968}, with a weak dependence on the temperature \citep{Fyfe1974}.

In another reaction brucite (\ce{Mg(OH)2}) is produced in addition to serpentine \citep{Martin1970}. This has not been found in large amounts in meteorites, however, it might have been thermally decomposed after its formation.
\begin{equation}
\ce{2Mg2SiO4 + 3H2O -> Mg3Si2O5(OH)4 + Mg(OH)2}
\label{eq:brucite}
\end{equation}

\noindent The actual rate of serpentinization depends on the composition of the rock and the ability of the liquid to transport magnesium and other elements during the process. 

Based on meteorite samples, it is generally estimated that these aqueous changes occur on a short time scale of about 100\,years, at low temperatures, but above the melting temperature of \ce{H2O}, supported by previous models \citep{Dufresne1962, Grimm1989, Zolensky1989}. \cite{Gobi2017} pointed out the importance of interfacial water. This type of water can exist at temperatures below the melting point of bulk "classical" water as a microscopic liquid film along mineral--H$_2$O interfaces at subzero temperatures, mainly due to van der Waals forces, enabling the serpentinization process to proceed. Based on the simulations they estimated that the reaction duration was comparable to time scales from previous studies.

When constructing a serpentinization heat model we have to take into account the amount of heat released, the heat consumption required to melt the ice, and the heat transfer as well.
There are several models for estimating heat production.
The simplest is the heat-balance model \citep{Lowell2002} which assumes the rock-water system is stationary and completely ignores thermal conduction.
The temperature dependence of reaction enthalpies and heat capacities has already been taken into account in the improved model by \citet{Allen2004}.
With these methods, only the increase in the temperature of the surrounding material can be estimated.
Heat loss during ice melting is only taken into account in dynamic models \citep{Cohen2000}.
In heat-balance models, convection of liquid water is considered, while in dynamic models, conduction in solid components and diffusion of molecules are considered to be the main modes of heat transfer.

The heat transfer significantly affects the total heat balance and can be influenced by a number of parameters, such as the porosity of the planetesimal.
The value of porosity in carbon-containing chondrites is around 20\% \citep{Consolmagno2008} while for smaller trans-Neptunian objects (TNOs) it can reach up to 60\% if they formed after the decay of most $^{26}$Al \citep{Bierson2019}. The porosity can affect melting and heat transfer due to the lack of continuity of the solid material as it reduces the conductive heat transfer but promotes convection by allowing fluid to migrate in the interconnected gaps.
Higher porosity can increase the rate of the reaction, as the effective surface area is larger, thus the contact surface of olivine and \ce{H2O} molecules also increases.
In this way, the amount of interfacial liquid water can also increase at temperatures below the melting point of water.
Furthermore, the interconnected porosity allows the liquid water to reach places where it has already run out, thus allowing the reaction to continue.

The known meteorite samples originated mainly from the main asteroid belt where the average densities of objects are much higher than in the outer Solar System ($\rho$\,$\geq$\,2\,g\,cm$^{-3}$ in contrast with the transneptunian region where even densities of $\rho$\,$\leq$\,1\,g\,cm$^{-3}$ are common).
In large Kuiper belt objects current estimates point to a common primordial (bulk) density of $\sim$1.8\,\gcc\, \citep{Barr2016,Grundy2019}, indicating a rock-to-ice ratio of $\sim$42:58 in volume, and $\sim$70:30 in mass. This is notably higher ice content than in the main belt, where e.g. the $\rho$\,=\,2.16\,\gcc\, density of (1)~Ceres \citep{Park2016} indicates of rock-to-ice mass ratio of $\sim$80:20, and the high $\rho$\,=\,3.5\,\gcc\, density of (4)~Vesta \citep{Russell2012} suggests a very low water ice content. If the primordial density in the Kuiper belt was really close to the value obtained from the bulk density of large objects ($\sim$1.8\,\gcc), small bodies in the D\,$<$\,500\,km range should have a notable macroporosity to have the observed bulk densities below $\sim$1\,\gcc\,\citep[see e.g.][for a recent evaluation]{Grundy2019}. The abundance of ice and the level of porosity indicate that aqueous alteration processes may have played an important role in the evolution of these objects, at least for a short time, early in the evolution of the Kuiper belt, when radiogenic decay provided enough heat for these reactions to start. 

In this paper, we present a revised model of the serpentinization process inside planetesimals. This model is incorporated into a general thermal evolution model.
The main outline of the model is presented in Sect.~\ref{sect:thermal_model}. We apply it to explore the role of this chemical process in the early evolution of planetesimals in the Kuiper belt in Sect.~\ref{sec:tno_example}. 
A detailed list of the equations used is given in the Appendix.

\section{Thermal evolution model of planetesimals \label{sect:thermal_model}}

Our serpentinization model is based on \cite{Gobi2017} which is an improved serpentinization model from heat balance \citep{Lowell2002, Allen2004} and dynamic models \citep{Cohen2000}. \cite{Gobi2017} used a kinetic approach to estimate the rate of serpentinization and follow its evolution.
Their model implemented interfacial water but included some simplifications and neglected several effects that did not cause a significant difference in the results for the smaller object ($\sim$15\,km radius), low water ice content and low porosity they studied.
In our cases, however, these neglected effects are important as we aim to consider larger planetesimals and icy bodies as well.  
In the description of the serpentinization model (in Sect.~\ref{sect:serp_model}. and \ref{sect:comp_results}.), we present which effects were taken into account in the current model as an improvement, compared with the previous model in which these parameters were simplified.

To create a complete heat development model, several other effects must be considered which are important in the early solar system: the decay of radionuclides (dominated by the short-lived $^{26}$Al), the accretion heat and the time of accretion with respect to the onset of radiogenic decay. The heat-generating capabilities of these heat sources were incorporated into the improved serpentinization model, and the heat loss/cooling of the planetesimals and the heat conduction were taken into account (in Sects.~\ref{sect:radio}. and \ref{sect:heat_transfer}.).

\subsection{Modelling the serpentinization process \label{sect:serp_model}}
In this part of the paper we focus solely on the algorithm which is able to follow the serpentinization process itself. We assume that the reaction described by Eq.~\ref{eq:pyrox} is the dominant process for serpentinite production in the planetesimals and we take only this reaction into account when we calculate the heat released, following \citet{Gobi2017}.
The planetesimal used in the model is made of seven components: silicate rocks are forsterite (Mg-rich olivine), enstatite (Mg pyroxen), hydrated rock (serpentinite), non-reactive solid material 
and \ce{H2O} in the three-phase state: liquid, solid and void space filled with \ce{H2O} vapor.
The melting point of \ce{H2O} (268\,K) is obtained from the properties of a saturated solution of \ce{MgSO4} \citep{Kargel1998}, which was used in model in earlier studies, too \citep{Cohen2000, Gobi2017}. While \ce{MgSO4} is quite common in some chondrites \citep{Zolensky1999} in the case of TNOs other solutes could also be considered. However, the effect of any salt or salt combination on the melting temperature of water is probably small.

\begin{figure}[!ht]
\begin{center}
\resizebox{6.0cm}{!}{\rotatebox{0}{\includegraphics{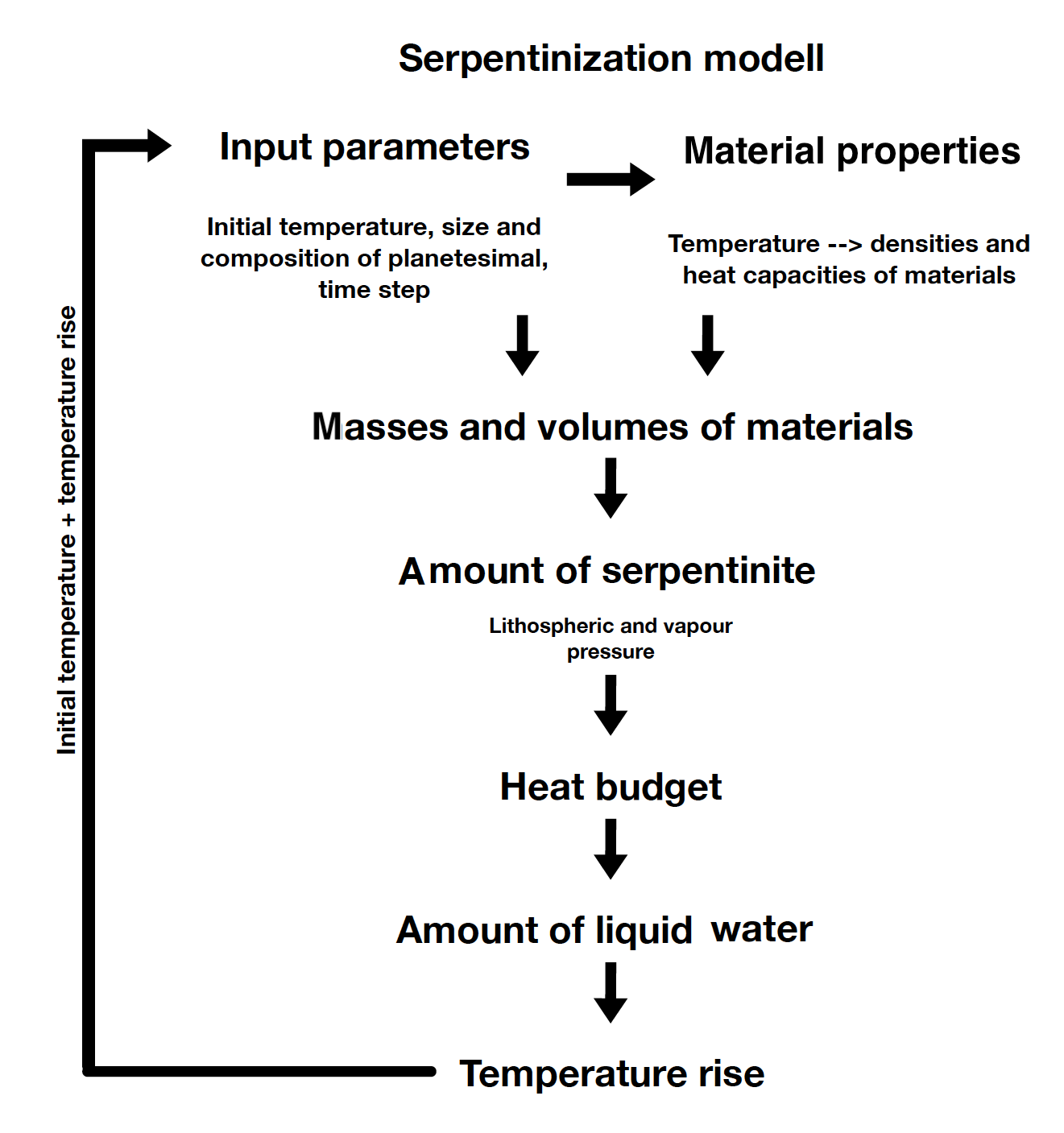}}}
\caption{The serpentinization model scheme.} \label{fig:model_scheme}
\end{center}
\end{figure}

To be comparable with the \citet{Gobi2017} model, the serpentinization algorithm itself is tested in a selected layer of a sphere, with the homogeneous and isotropic distribution of the components within that layer. Due to the simple forward Euler scheme, the model remains sensitive to the choice of the time step. We explore this in Sect.~\ref{sect:comp_results} to find the optimal (largest allowable) time step for our calculations.
Apart from the improvements discussed below, this is the same scheme as was used in \citet{Gobi2017} (see the model scheme in Fig \ref{fig:model_scheme}.).

\paragraph{Heat capacity}
Heat properties of materials and the average values of the physical properties of planetesimal depend on the temperature and composition (see in Sect.~\ref{app_prop}). We calculated the heat capacity of \ce{H2O} in a different, more accurate way than in the base model.
After fixing it, the final heat production was higher by a few Kelvin. However, the difference remained within the uncertainty range, even at longer timescales. 

\paragraph{Lithospheric pressure}
Several quantities depend on the total pressure which is calculated as the sum of lithospheric and vapor pressures (see in Sect.~\ref{app:am_serp}). 

\begin{figure}[!ht]
\begin{center}
\resizebox{8.0cm}{!}{\rotatebox{0}{\includegraphics{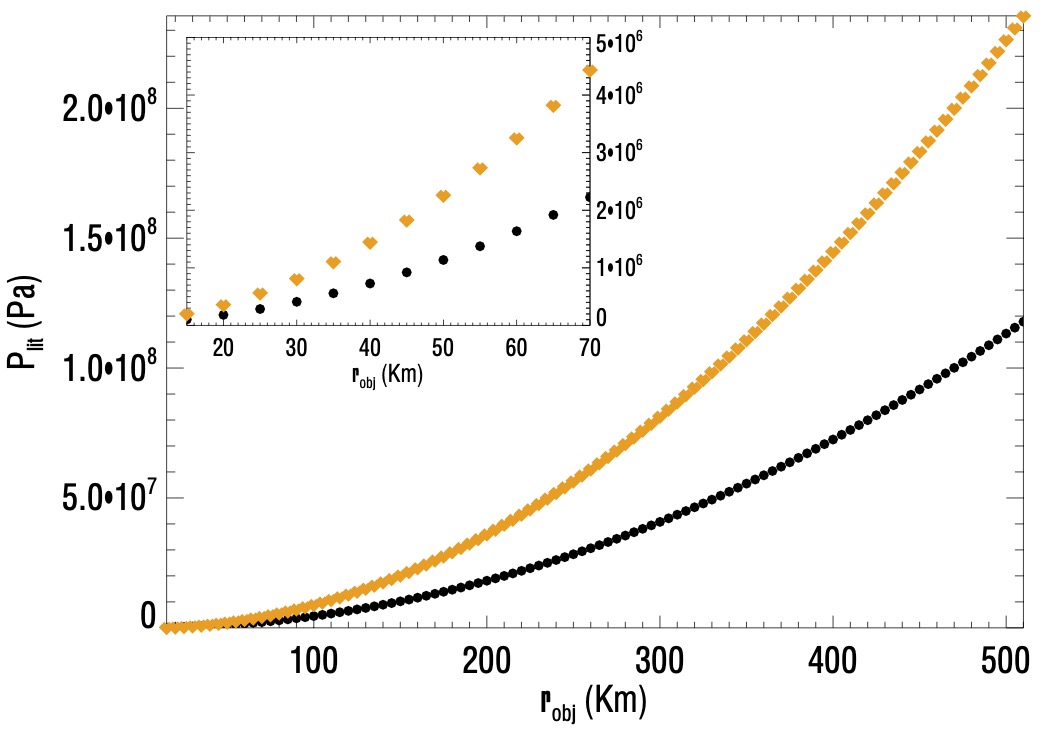}}}
\caption{Lithospheric pressure in the center of the test object versus its radius.
Orange diamonds represent the simplified calculation of lithosphere pressure \citep{Gobi2017} while the black curve shows the results of the calculations using our more accurate model.} \label{fig:Plit}
\end{center}
\end{figure}

\citet{Gobi2017} used a simplification to determine the lithospheric pressure due to the small size (R\,=\,15\,km) of their main test objects, as the contribution of the lithospheric pressure to total pressure is insignificant for small objects. This was replaced by Eq.~\ref{eq:P_lit} which gives reliable results for larger objects, too. Using this method the reaction rate becomes lower which makes a more significant difference for larger objects.
 We compare our results with the \citet{Gobi2017} litospheric pressure values in Fig.~\ref{fig:Plit}.
\paragraph{Interfacial liquid water}
To obtain the heat gain we need to know the exact amount of liquid water and the latent heat of water vaporization (Sect.~\ref{app:heat_gain}).
The latent heat was calculated in a way different from the base model and it caused a few tenths of Kelvin difference in the final temperature, with a very small amount of vapor formed under typical conditions.
In those cases when the temperature is lower than the melting point of \ce{H2O}, ice is present instead of liquid water which would not allow serpentinization to take place.
However, \cite{Gobi2017} pointed out that the presence of microscopic scale interfacial water is possible on the surface of olivine grains at low temperatures making serpentinization possible, and it may gradually melt the icy surrounding of silicate particles due to the heat produced in this exothermic reaction.
The model of \cite{Gobi2017} allowed the formation of more interfacial water than it would have actually been possible at the given temperature.
They did not take into account the rate of serpentinization, although this effect is significant in the temperature range below the melting point of bulk ice.
In our model we consider four values for the amount of water:
i) the actual amount of ice in the layer examined, 
ii) the maximum value of interfacial water that can be formed,
iii) the amount of water that is able to react in a specific step, and
iv) the amount of ice that can be melted by the serpentinization heat (see Eq.~\ref{eq:meltice}). The {\it minimum} of these four values determines the actual amount of interfacial water which can be formed.  

In the \cite{Gobi2017} paper the improper handling of the interfacial water also caused an issue in selecting the proper simulation time step: if  larger time steps for the same timespan were used, the calculated initial temperatures required for the same serpentinization level were lower and the final temperatures higher, causing a greater overall heat production in the simulations.
The reason behind this phenomenon was that in the case of decreased time steps the serpentinization reaction started later in time.
By incorporating the new condition in the melting ice calculation, the simulations did not produce more interfacial water any more than it was possible at the given temperature, and there was no difference in the amount of heat produced as a function of time scale at the same time.

\subsection{Serpentinization model results compared with previous results  \label{sect:comp_results}}

To demonstrate the capabilities of our improved algorithm to follow the serpentizination process we used the same setup as \citet{Gobi2017}, but with the corrections listed above.
We consider the deepest 100\,m radius, without taking other heat sources and heat/material transfer into account.
In each case the amount of non-reacting material of the test object was constant, 14\% of the volume of planetesimals. We investigated how the different parameters of the model can influence the outcome compared with the \citet{Gobi2017} results.

\paragraph{Initial temperature}

\begin{figure}[!ht]
\begin{center}
\resizebox{8.0cm}{!}{\rotatebox{0}{\includegraphics{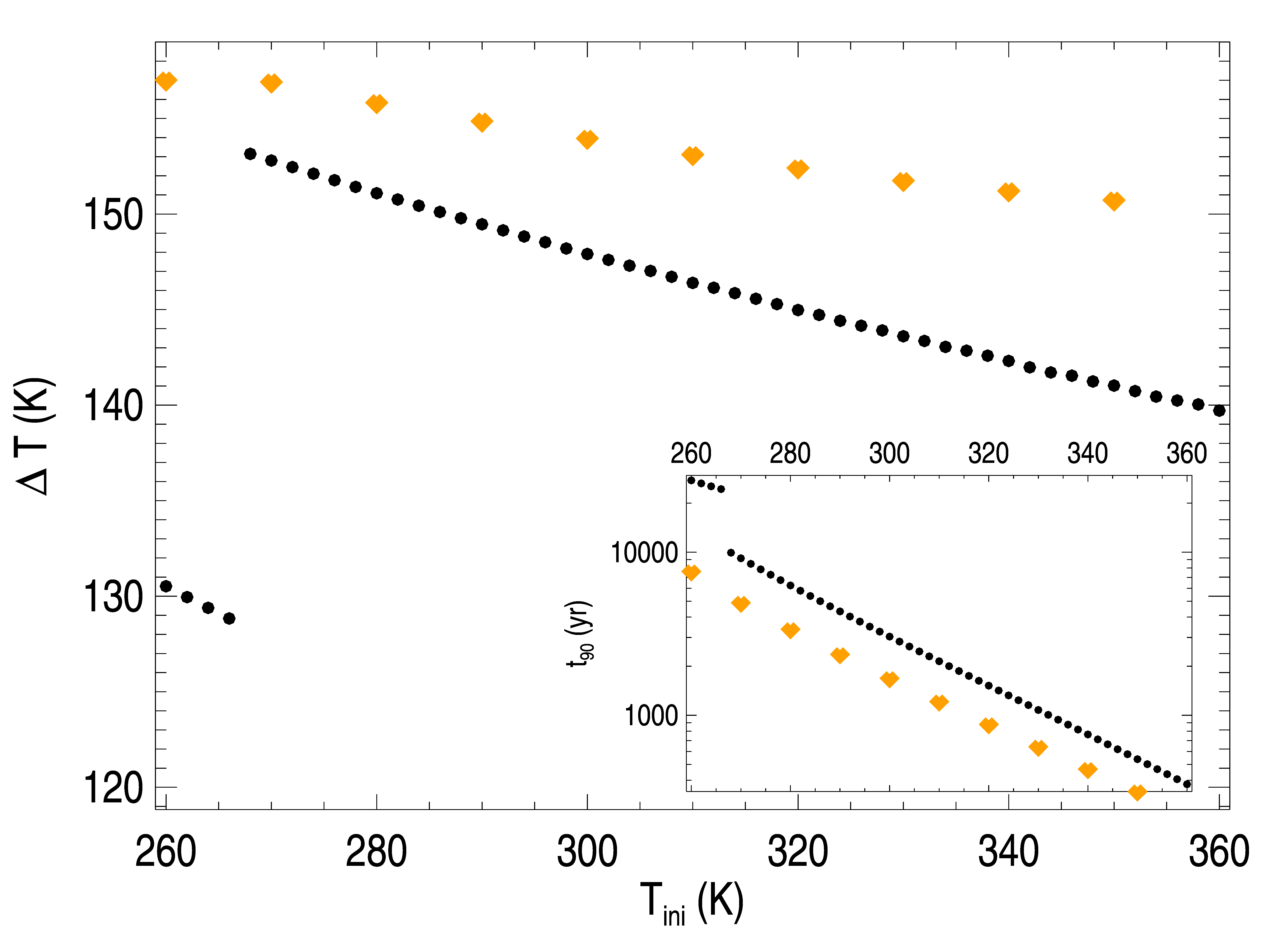}}}
\caption{Serpentinization time (t$_{90}$) and temperature increase ($\Delta$T) during the serpentinization process. 
Orange diamonds represent the results by \citet{Gobi2017} for an object with 15\,km radius and the black dots represents our improved model for a same object, using a time step of 0.1\,year. Both test objects have 16\% porosity and 1:2 olivine-to-water ratio. Here we examined the innermost 100\,m radius of the object, to be comparable with the previous results.} \label{fig:time}
\end{center}
\end{figure}

The output of any evolutionary model that includes chemical reactions like serpentinization is very sensitive to the choice of initial temperature.
We examined the temperature increase ($\Delta$T) during the reaction as well as the time needed to consume 90\% of the reagent material during the serpentinization reaction ($t_{90}$)  at different initial temperature (T$_{ini}$) values (Fig.~\ref{fig:time}). The temperature dependence of chemical reactions is known in general: the higher the initial temperature, the faster the reaction. The dependence of the temperature increase on the initial temperature is a result of the temperature dependence of the heat capacity of the constituent minerals (equations for calculating the heat capacities can be found in Sect.~\ref{app_prop}).

The serpentinization time t$_{90}$ is significantly longer than in the previous studies (Fig.~\ref{fig:time}), and the serpentinization rate is higher towards higher initial temperatures. Mainly due to the accurate calculation of the lithospheric pressure and the consideration of the maximum value of sub-freezing interfacial water t$_{90}$ increased.
In the case of a planetesimal with a radius of 15\,km, the process can be up to 3-4 times longer than in the previous calculations, depending on the initial temperature. 

This is due to the fact that only a small amount of interfacial liquid water is present in the system below the freezing point and its production depends on the amount of water reacted, thus the process is slow. When the temperature reaches the melting point the heat production is entirely used to melt the ice. Above the melting point, the reaction speeds up due to the accessibility of large amounts of liquid water. Starting the reaction below the melting temperature, most of the heat is consumed by melting the ice and this results in an overall smaller temperature increase.

\paragraph{Importance of microscopic liquid water}

Interfacial liquid water is an important component as it promotes the progress of the reaction in the early stage when the temperature is below the melting point of ice. Compared with previous results, it can be seen that in the sub-freezing initial temperature range, the process is slower and the heat production rate is also lower (Fig.~\ref{fig:time}). The reaction rate increases after the melting point.

\paragraph{Size of planetesimals}

The reaction rate increases with increasing pressure and it depends both on the size of the test object through the lithospheric pressure and the vapor pressure (Sect.~\ref{app:am_serp}).
The pressure dependence of serpentinization was studied previously \citep{Martin1970, Wegner1983, Jones2006, Cohen2000}, and it was found to be nearly linear in the range of 1-200\,MPa which corresponds to the pressure expected in the size range in our investigation.

\begin{figure}[!ht]
\begin{center}
\resizebox{8.0cm}{!}{\rotatebox{0}{\includegraphics{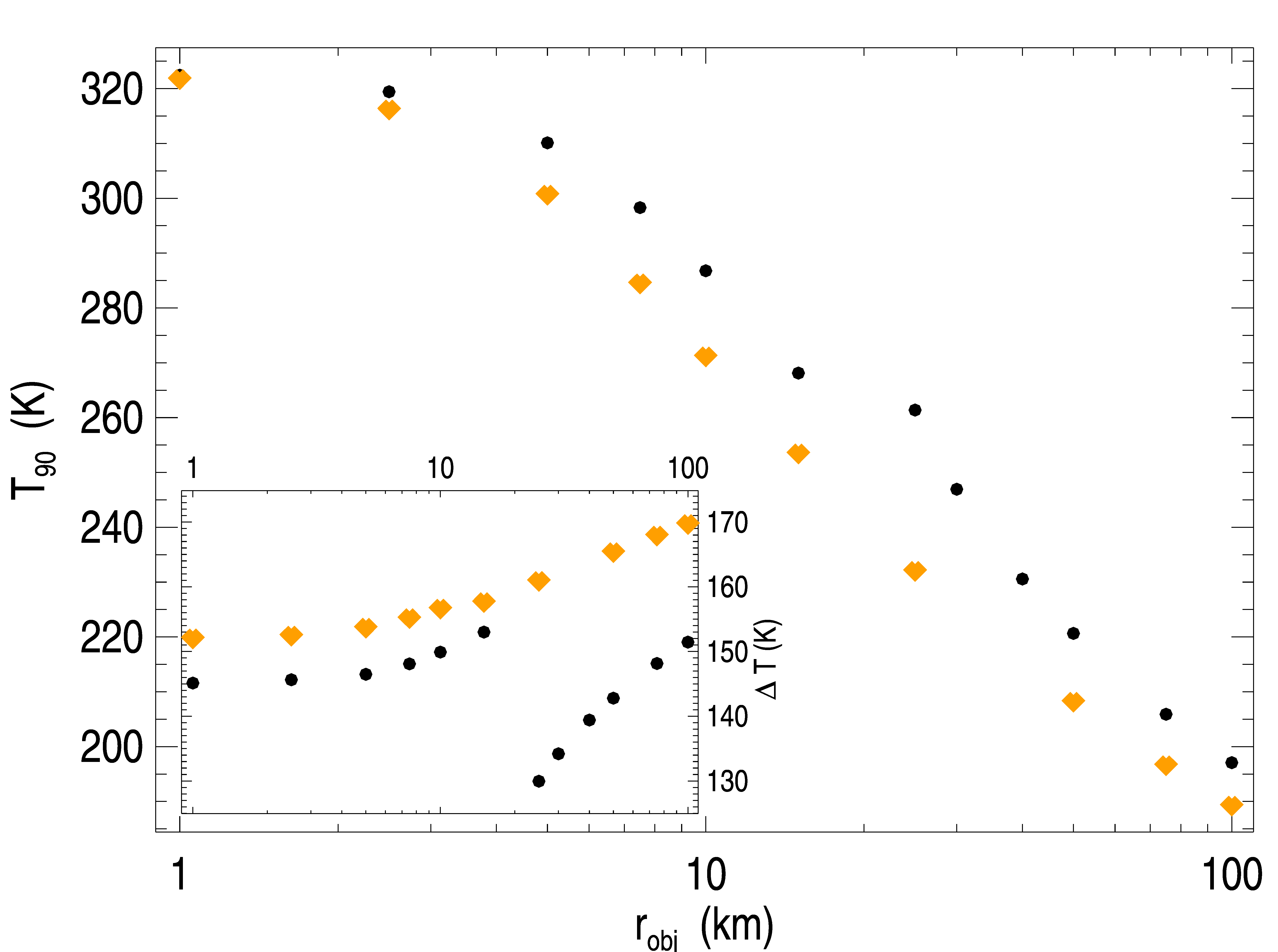}}}
\caption{The initial temperature (T$_{90}$) and temperature increase ($\Delta$T) for different sizes of planetesimals (r$_{obj}$) 
in $t_{90}$, using 1:2 olivine to water ratio. The filled circles and the orange diamonds mark the results from this present work and from the previous study by \citep{Gobi2017}, respectively.}\label{fig:size}
\end{center}
\end{figure}

The initial temperature required for serpentinization to consume 90\% of the materials in 10,000 years (T$_{90}$) reaches values similar to those determined in the previous work \citep{Gobi2017}, the deviation is only a few K (see Fig.~\ref{fig:size}). The largest deviation occurs at the size/pressure value where the initial temperature falls below the freezing point.

There is an increasing trend in temperature change ($\Delta$T) with increasing size (see Fig.~\ref{fig:size}) due to heat capacities which are lower due to the lower initial temperature (T$_{90}$). When the required initial temperature falls below the freezing point the heat production rate decreases significantly as some of the heat produced is used to melt ice. From this point on, the $\Delta$T again shows an increasing trend as a function of object size.

\paragraph{Olivine to \ce{H2O} ratio and initial porosity}

The component ratio of olivine-to-water is one of the most important initial parameters of the serpentinization reaction which significantly influences both the course and the outcome of the reaction. In the case of higher olivine-to-water ratios, the reaction can be faster than in the cases of higher water content \citep[see e.g. Fig.~1. in ][]{Gobi2017}. For each olivine-to-water ratio value, a higher T$_{90}$ is needed (see e.g. in Fig.~\ref{fig:size}), but the difference is only a few K.
The rate of heat production is closely related to heat capacity which increases with water content as the heat capacity of water is higher than that of rocky components. This slows down the reaction towards higher water content.

\paragraph{Time step}

\begin{figure}[!ht]
\begin{center}
\resizebox{8.0cm}{!}{\rotatebox{0}{\includegraphics{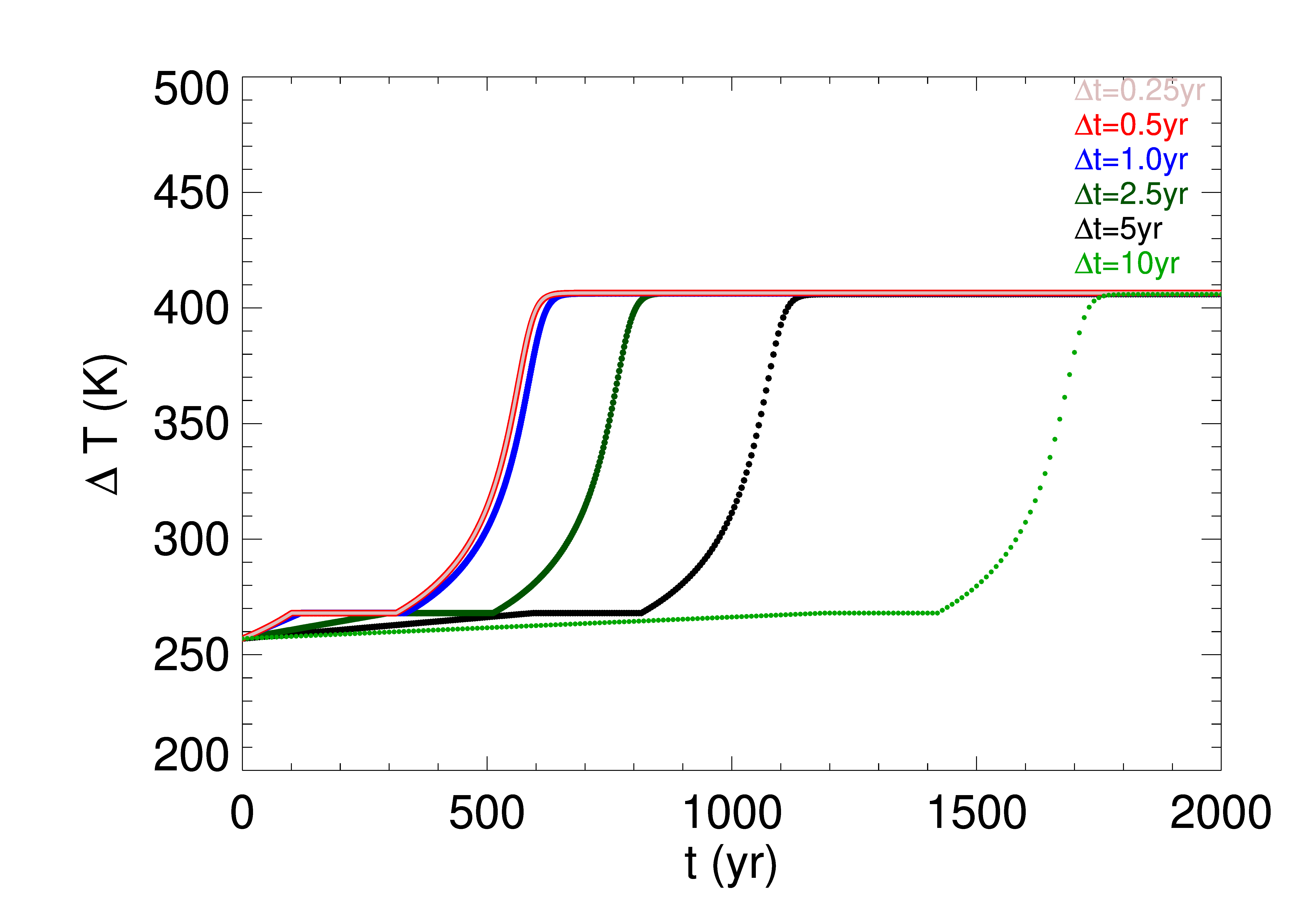}}}
\resizebox{8.0cm}{!}{\rotatebox{0}{\includegraphics{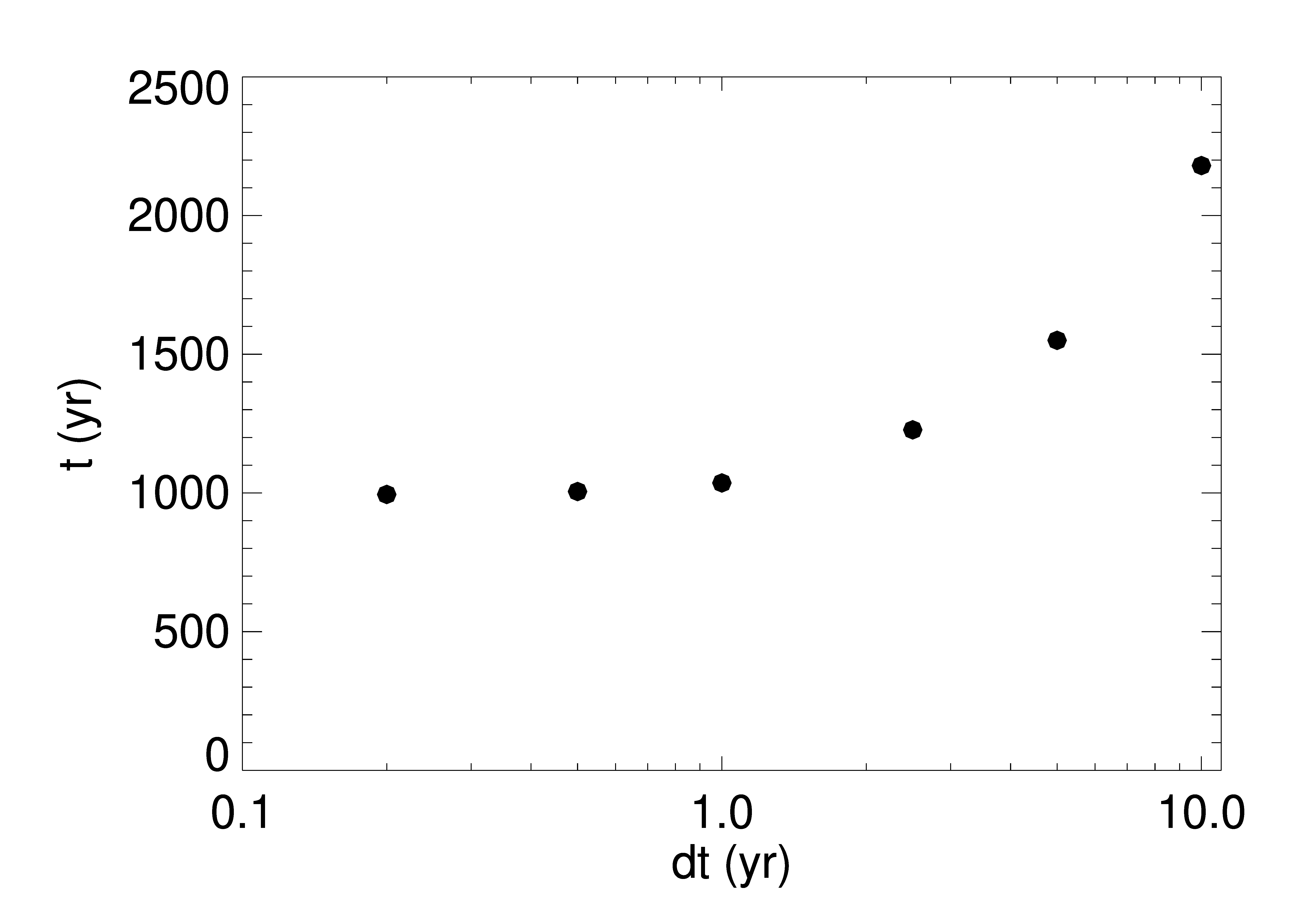}}}
\caption{Upper panel: temperature increase due to serpentinization as a function of the time required (when one of the reagent materials is 100\% consumed). Different colors show different model time steps. Bottom panel, the time required for chemical reaction as a function of the model time steps. \label{fig:time_steps}}
\end{center}
\end{figure}

As we use a simple forward-Euler scheme in our model, the results are expected to be sensitive to the time step applied. Choosing a large timestep ($\Delta$t\,=\,10--100\,yr, or larger) results in considerable instability in the calculations, i.e. the final results (e.g. T$_{90}$ or $\Delta$T we investigated above) do not show a consistent trend, and depend strongly on the time step chosen. This annoying effect disappears when the timestep is decreased, and the calculations become stable for $\Delta$t\,$\lesssim$\,1\,yr (see Fig.~\ref{fig:time_steps}). To avoid this problem we used a timestep of $\Delta$t\,=\,0.5\,yr. Reducing the timestep further did not cause a considerable change in the final results.

\subsection{Decay of radionuclides \label{sect:radio}}

\begin{figure}[!ht]
\begin{center}
\resizebox{8.0cm}{!}{\rotatebox{0}{\includegraphics{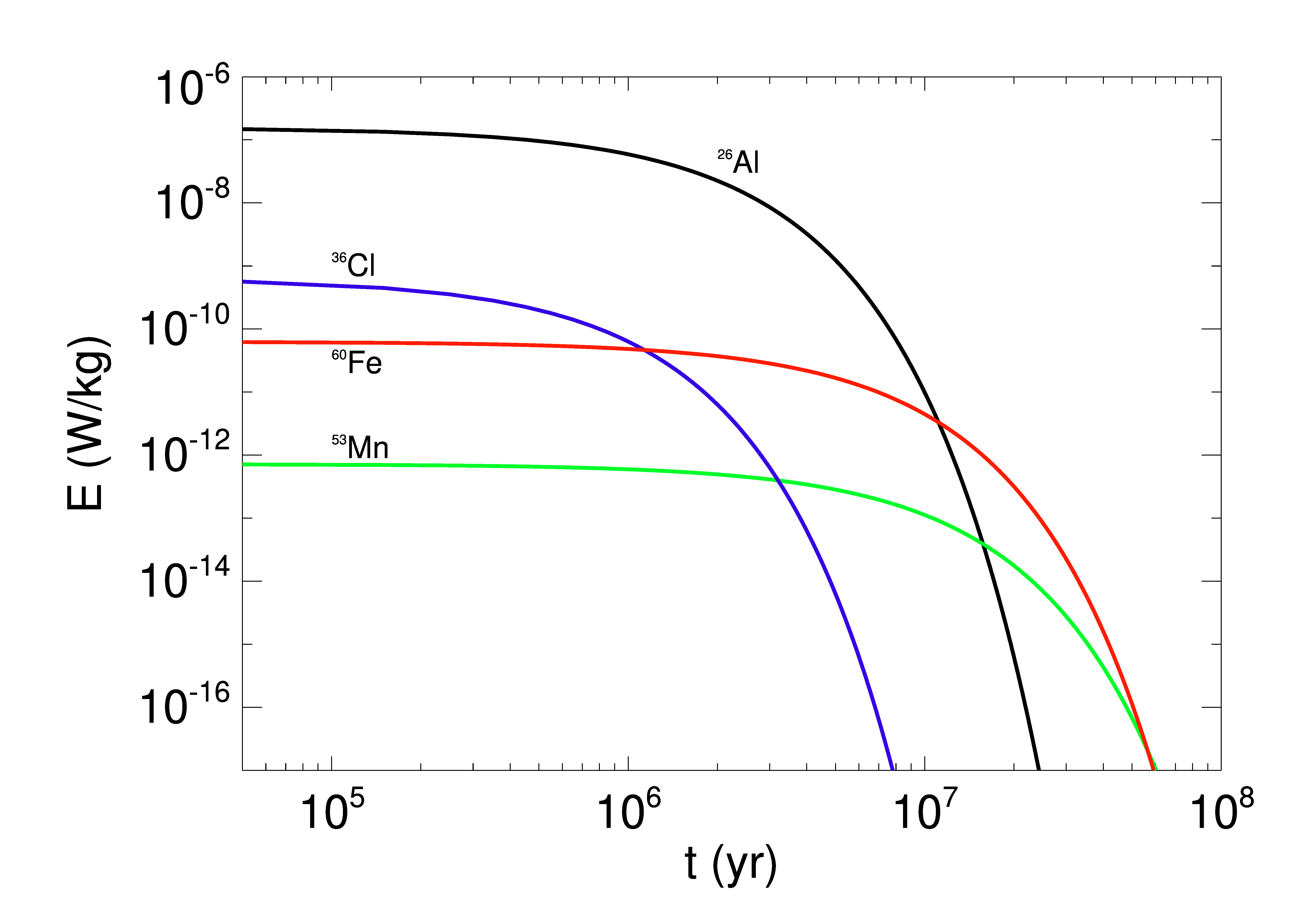}}}
\caption{Total energy output as a function of time for some short-lived radionuclides. Time scale starting 50,000 years after the time when the radionuclides start to decay. } \label{fig:rad_energy}
\end{center}
\end{figure}

The main source of heat within planetesimals is the radioactive decay of both short- and long-lived radionuclides. In Fig.~\ref{fig:rad_energy} we plot the total energy output as a function of time for some short-lived radionuclides. Because serpentinization is a rapid process, only a few 10,000 years, we investigated the early stages of the heat development of planetesimals. During this period, as shown in Fig.~\ref{fig:rad_energy}, the most significant portion of the heat production from radioactive decay was provided by $^{26}$Al \citep{Lugaro2018}, and only this isotope was considered in our thermal evolution model as a heat source, with the following parameters: half-life is t$_{1/2}=0.717$\,Myr, total energy is TE$=5.07 \times 10^6$ \,J/kg, and we assume that $^{26}Al$ was homogeneously distributed in the early solar system \citep{Lugaro2018}. 

\subsection{Heat transfer} \label{sect:heat_transfer}

There are two major ways of heat transfer inside a planetary body: thermal conduction and convection. Both of them were considered, and the thermal radiation from the surface was also included in the thermal evolution model.

\paragraph{Thermal radiation}
The power output was calculated from the Stefan-Boltzmann law using a constant emissivity factor ($\epsilon =0.9$), and assuming an ambient temperature of T$_{amb}$ = 50\,K  which is a typical surface temperature in the outer Solar System (Eq.~\ref{eq:sug}).

\paragraph{Thermal conduction}

The heat conduction rate was estimated with the Eq.~\ref{eq:cond}, in all boundaries of all layers following \cite{Hussmann2006}.

\paragraph{Thermal convection}

If the temperatures within the layers are sufficiently high and water is present in a liquid state convective motions can occur in the porous media \citep{Hewitt2014}. The requirement for starting convection is that the Rayleigh number (Ra) exceeds a specific value. Ra is calculated in a porous material as:
\begin{gather}
\begin{split}
Ra & = \frac{\rho\,\beta_0\,\Delta T\,k_0\,g\,dr}{\eta\,\alpha} \\
\end{split}
\end{gather}
\noindent where $\rho$ is the density in the layer, $\beta_0$ is the thermal expansion coefficient and we used $\beta_0\,=\,10^{-3}$ as a safe upper limit for any of the possible constituents, $\Delta$T is the temperature difference across distance $dr$, which is the thickness of the layer, $k_0$ is the permeability, and we used two values: $10^{-12}\,m^2$ following \cite{Cohen2000} and $10^{-7}\,m^2$ following \cite{Jacob1972}. The results were the same because the Ra did not reach the critical value required for the start of convection with any of the permeability values. $g$ is the local gravitational acceleration, $\alpha = k/(\rho \cdot c_p)$ is the thermal diffusivity ($k$ is thermal conductivity and $c_p$ is specific heat capacity) and $\eta$ is the  dynamic viscosity of the fluid:
\begin{gather}\label{eq:alpha_eta}
    \eta(T) = \eta_0 \, exp(25\times(T_{melt}/T-1))
\end{gather}
\noindent where $\eta_0 = 10^{13}$\,Pa\,s is the melting point viscosity \citep{Hussmann2006} 
and T$_{melt}$ is the melting point. A critical Rayleigh number of 1000  was used for the onset of convection following \cite{Hussmann2006}. However this critical value can be as low as 40 \citep{Nield1999} in a porous medium. In our simulations these Ra values have not been reached at T\,$\lesssim$\,T$_{melt}$. 

The main purpose of our present work is to follow the evolution of the serpentinization process, and to determine the conditions and onset timescales this process can operate at. As we show it below, before and during the active serpentinization stage the temperature in all of our simulations is close to or below the melting point of ice, and the heat transfer is governed by these 'icy' conditions. During serpentinization the melted water is quickly consumed by the chemical reaction, preventing convection (which would be caused by the otherwise significantly decreased viscosity of liquid water). After the end of the serpentinization stage radiogenic decay can still provide extra heat to further melt the ice. In these conditions the low viscosity of liquid water may lead to convection, generating a faster heat transfer within the inner layers and may also lead to rearrangements in the structure of these layers. The consideration of this process, as well as e.g that of the diffusion of water into solid grains, are beyond the scope of this paper.

\section{Application to transneptunian objects} \label{sec:tno_example}

\subsection{Main model outline}
Serpentine is a primary candidate to explain the reflectance spectra of transneptunian objects \citep[e.g.][]{Protopapa2009}, and it has also been considered as primordial material in impact simulations aimed to explain the formation of transneptunian binary system \citep[e.g. the Pluto-Charon system;][]{Canup2005}.
Therefore it is an intriguing question how serpentine can form under the conditions in the transneptunian region where objects have typically high ice contents and possibly high porosity, and how far it can contribute to the heat budget and thermal evolution of these planetesimals.

We consider a spherical body with spherical symmetry (all variables depend on the radius only), with a size-dependent number of layers. We assumed a homogeneous and isotropic composition at the start, and a common T$_{ini}$ in the whole planetesimal. The effect of deviations from this latter assumption is investigated in Sect.~\ref{sect:serp_eff}. Heat production by radiogenic decay and heat transfer is considered as described in Sects.~\ref{sect:radio} and \ref{sect:heat_transfer}.

\subsection{Initial parameters}\label{subsec:in}

\paragraph{Composition}
The low density of TNOs is due partly to their composition, as the \ce{H2O} content of these distant objects is significantly higher than those in the inner Solar system, but it is also due to their higher porosity.  In the case of smaller TNOs, when the radius does not exceed $\sim$150\,km, the porosity can reach 60\% \citep{Bierson2019}. Such a high porosity is only possible if they formed after the decay of $^{26}$Al to maintain their high porosity as there is no internal transformation by heat at this stage.
In the absence of accurate knowledge of the internal composition, a simplified composition was used, which has been used in previous serpentinization studies \citep[See in][and in the \ref{app:serp}]{Gobi2017,Cohen2000}.

\begin{figure}[!ht]
\begin{center}
\resizebox{7.50cm}{!}{\rotatebox{0}{\includegraphics{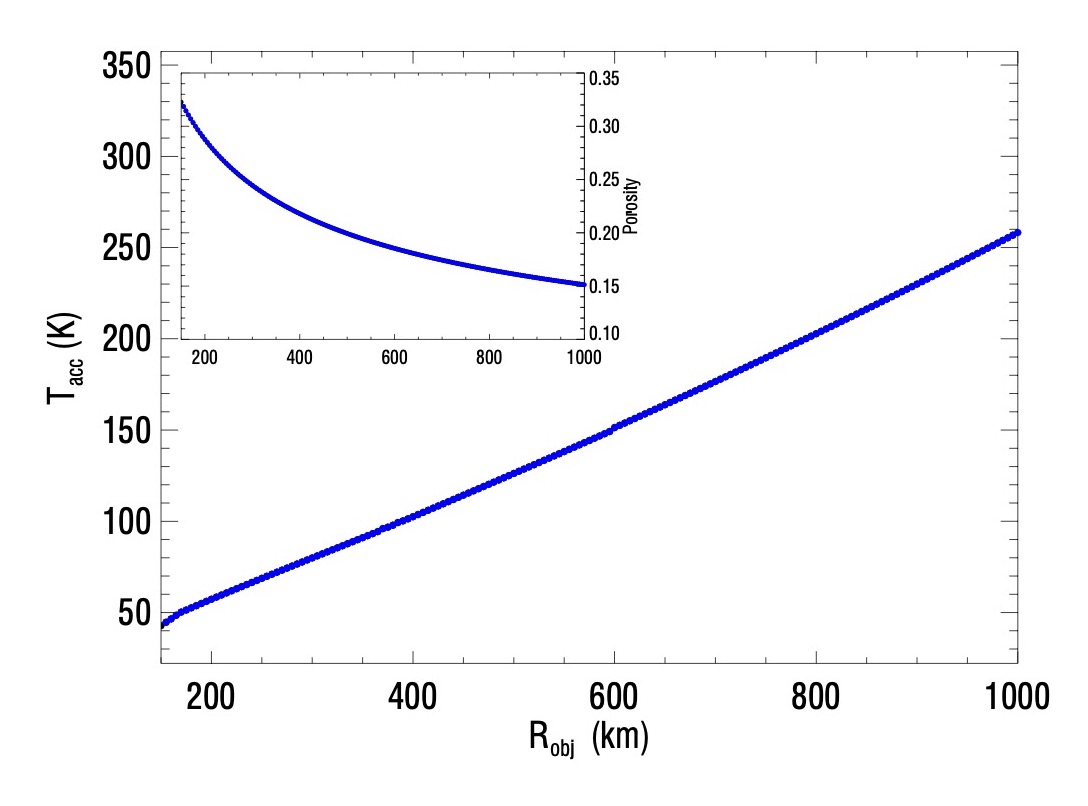}}}
\caption{The accretion heat product (T$_{acc}$) and porosity versus the size of the planetesimal (r$_{obj}$), for the case of 42:58 rock/water ratio with size dependent porosity.}\label{fig:tno_acc}
\end{center}
\end{figure}

\paragraph{Porosity}
Porosity is a very important factor in this model, the initial temperature (T$_{ini}$) and the serpentinization time 
also strongly dependent on it. To estimate the porosity, we used the calculations from \cite{Yasui2009} where they determine a size/pressure-dependent porosity for small icy bodies. The equation defined for the range Regime 3 (P\,$>$\,2\,MPa) was applied due to the pressure conditions in our objects:
\begin{gather}\label{eq:porosity}
\phi = a_{3}P^{b_3}
\end{gather}
\noindent where P is the lithospheric pressure (in MPa) in a middle layer in the measured bodies and $a_3$ and $b_3$ are constants. We used the approximate value of $a_3 = 0.5$ and $b_3 = -0.2$ \citep[see][]{Yasui2009}. With these calculations, the porosity values are between 13 and 34\% in the examined size range for the 42:58 rock/water ratio specified earlier (see Fig.~\ref{fig:tno_acc}) 

\paragraph{Temperature}
In the early solar system the initial temperature of the objects was determined by the size, composition, accretion heat, and the heat produced by the decay of short-lived radioactive nuclei. The heating from solar irradiation may be important closer to the Sun but it is negligible in the outer regions. 
In our model, the initial temperature was taken to be the heat from the accretion \citep{Hanks1969}:
\begin{gather}\label{eq:tacc}
T_{acc} = \frac{3}{5}\frac{GM}{RC_p}
\end{gather}
\noindent where $G$ is the gravitational constant, $C_p$ is the average heat capacity, $M$ is the mass and $R$ is the radius of the planetesimal (see Fig.~\ref{fig:tno_acc}.). At t\,=\,0 the planetesimal has a homogeneous temperature distribution with T$_{acc}$.
Note that as $C_p$ is temperature-dependent Eq.~\ref{eq:tacc} is an implicit equation for T$_{acc}$, and it is solved iteratively; it also determines the shape of the T$_{acc}$ curve presented in Fig.~\ref{fig:tno_acc}.

In a next step we tested the sensitivity of the radioactive decay heat product on the time step and layer thickness. In these tests we used the rock/water ratio of 42:58 that corresponds to an olivine to water ratio of 0.12. Radiogenic heat production was not very sensitive to the variation to time step and layer thickness. 

In all simulation configurations, we used a non-reactive rock content of 14\% with a density of 3630 kg\,m$^{-3}$ like the earlier studies \citep{Gobi2017}. In each case, the olivine water ratio was taken as 0.12, which corresponds to the assumed rock-water ratio of 42:58 in TNOs. The calculations are performed with the size-dependent porosity (Eq.~\ref{eq:porosity}.), the initial temperature is calculated from the accretion heat (Eq.~\ref{eq:tacc}.), and we choose a time step of 0.5\,yr and a layer thickness of 20\,km. The starting time t\,=\,0 was considered to be the end of the accretion.
\subsection{The effect of serpentinization in the thermal history of trans-Neptunian objects \label{sect:serp_eff}}

Objects with a radius below 150\,km are expected to have high porosity (up to 60\%) and the initial (accretion) temperature would remain very low ($<$\,42\,K), as indicated by our calculations. These objects may have formed after the depletion of $^{26}$Al \citep[as discussed in][]{Bierson2019}, and the serpentinization reaction could likely not produce a significant amount of heat. Therefore we did not consider these objects in our further calculations. Very large objects (R\,$>$\,1000\,km) were also excluded from the study size range because they are likely formed by multiple accretion events and undergo fast chemical differentiation early in their evolution.

\begin{figure}[!ht]
\begin{center}
\resizebox{8.0cm}{!}{\rotatebox{0}{\includegraphics{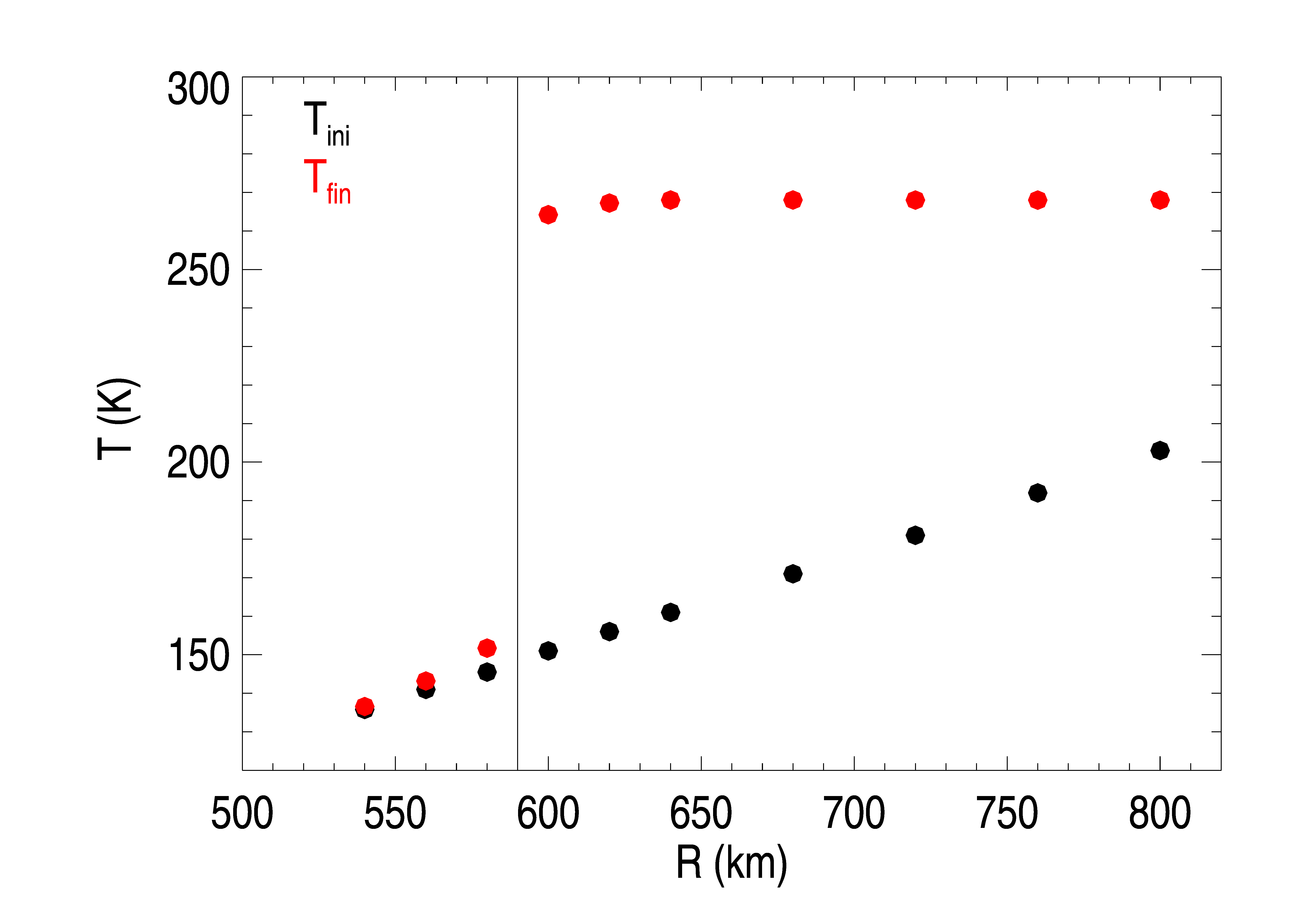}}}
\caption{Heat production by serpentinization without radioactive decay during the first 100,000 years, for different object sizes, which corresponds to different porosities and initial temperatures as discussed in Sect\ref{subsec:in}. The black symbols represent the initial temperature (due to accretion heat) and the red symbols show the final temperature in the center of the planetesimals.}\label{fig:serp_start}
\end{center}
\end{figure}

\begin{figure}[!ht]
\begin{center}
\resizebox{8.0cm}{!}{\rotatebox{0}{\includegraphics{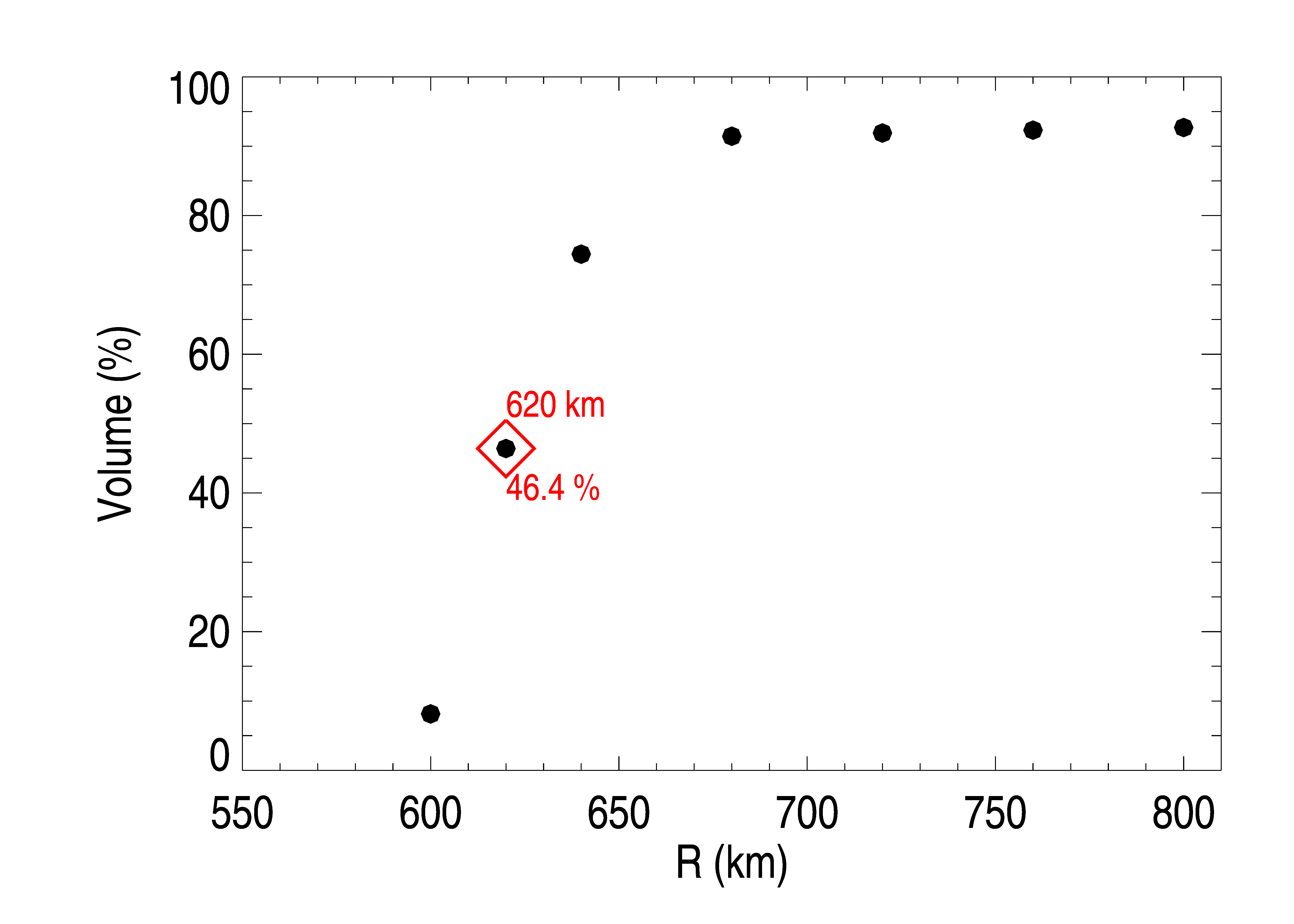}}}
\resizebox{8.0cm}{!}{\rotatebox{0}{\includegraphics{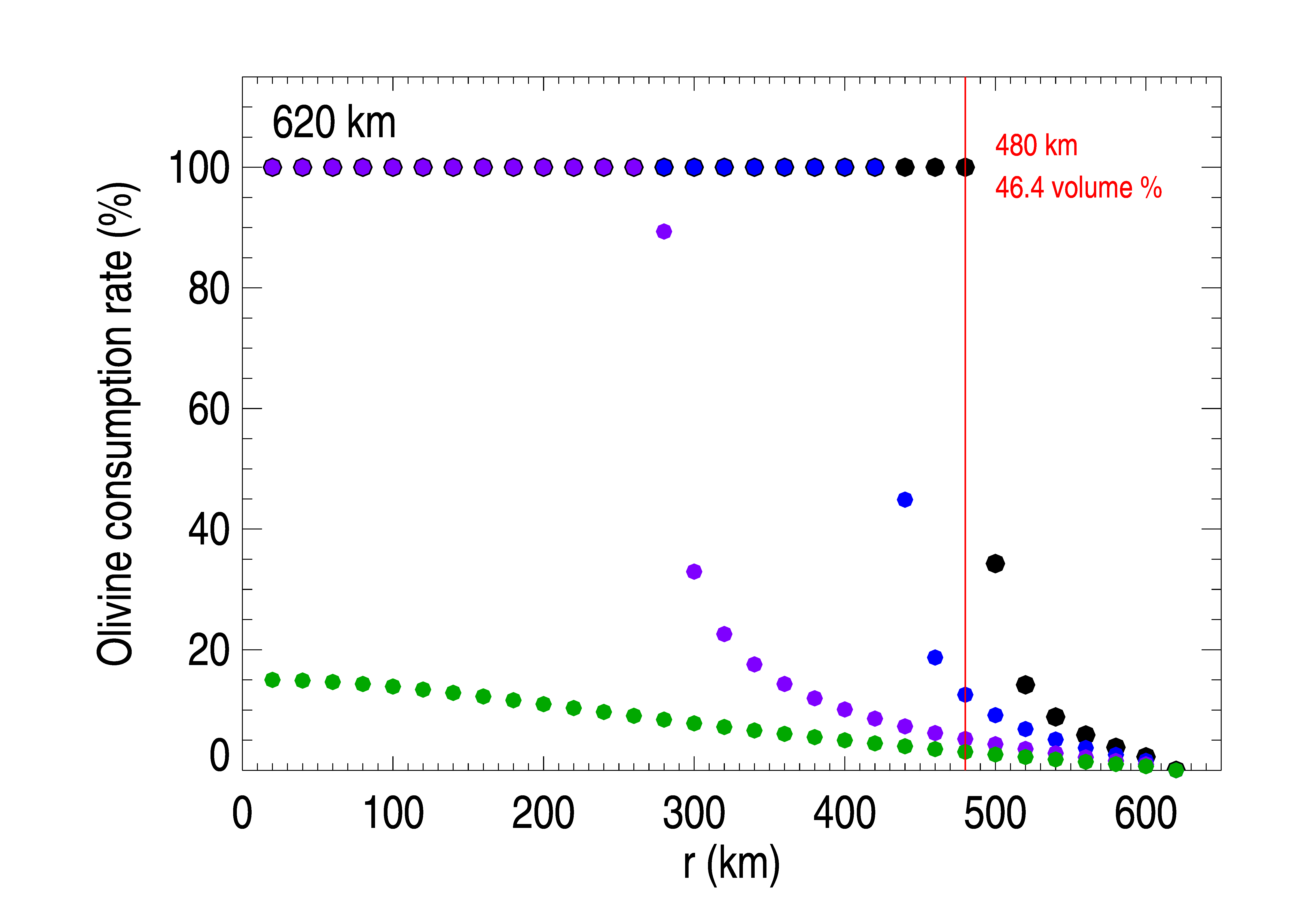}}}
\caption{Upper panel: ratio of internal volume in which 90\% olivine has been consumed in the first 100,000 years versus the objects size. Bottom panel: Progress of the serpentinization reaction, presented with the olivine consumption as a function of depth at different times, for a planetesimal of R\,=\,620\,km. The different colors mark 35,000 (green); 50,000 (blue); 75,000 (blue) and 100,000 (black) years.}\label{fig:serp_layers}
\end{center}
\end{figure}

We examined the conditions under which serpentinization can produce significant heat over a few tens of thousands of years in the size range of $150\le R\le 1000$\,km. We found that for objects smaller than 600\,km, the initial temperature from the accretion will be so low that the reaction cannot start or will be very slow, even in the core, {\it in the absence of radiogenic decay}. This appears in Fig.~\ref{fig:serp_start} as a 'jump' in temperature at this specific size. In larger objects, the process is already able to produce a notable amount of heat without the contribution of radioactive elements. Above an initial temperature of $\sim$150\,K, assuming the previously determined material composition and pressure-dependent porosity (Fig.~\ref{fig:tno_acc}) the reaction rate increases significantly (Fig.~\ref{fig:serp_start}). 
As the size increases, both the initial temperature and the internal pressure increase, resulting in an increase in the rate of serpentinization. This results in the extension of the central region where the serpentinization process can produce significant heat in a few tens of thousand years. The upper panel in Fig.~\ref{fig:serp_layers} shows the ratio of the internal volume in which 90\% olivine has been consumed in the first 100,000 years for a planetesimal with a specific size. For radii R\,$\ge$\,650\,km the serpentinized zone extends almost to the surface while it remains in the core below R\,=\,600\,km, in agreement with the slow reaction rate indicated by the small temperature increase seen in Fig.\ref{fig:serp_start}. The bottom panel in Fig.~\ref{fig:serp_layers} shows an example of how serpentinization proceeds inside an object of R\,=\,620\,km in the first 100,000 years. As shown also in the upper panel, a maximum of 480\,km radius of the serpentinized region is reached corresponding to a maximum volume ratio of 46\%. This R\,$\approx$\,600\,km critical size limit for the efficient progress of the serpentinization process was obtained without the consideration of radiogenic decay, and it already indicates that serpentinization can be an efficient process for the large objects (the dwarf planets of the trans-Neptunian region) even if they formed after 10$^7$\,yr or later, when the heat produced by radiogenic decay is significantly lower (see Fig.~\ref{fig:rad_energy}). This also suggests that the size limit of efficient serpentinization is smaller when an additional heat source --  radogenic decay -- is considered.


\begin{figure*}[!ht]
\begin{center}
\hbox{
\resizebox{7.8cm}{!}{\rotatebox{0}{\includegraphics{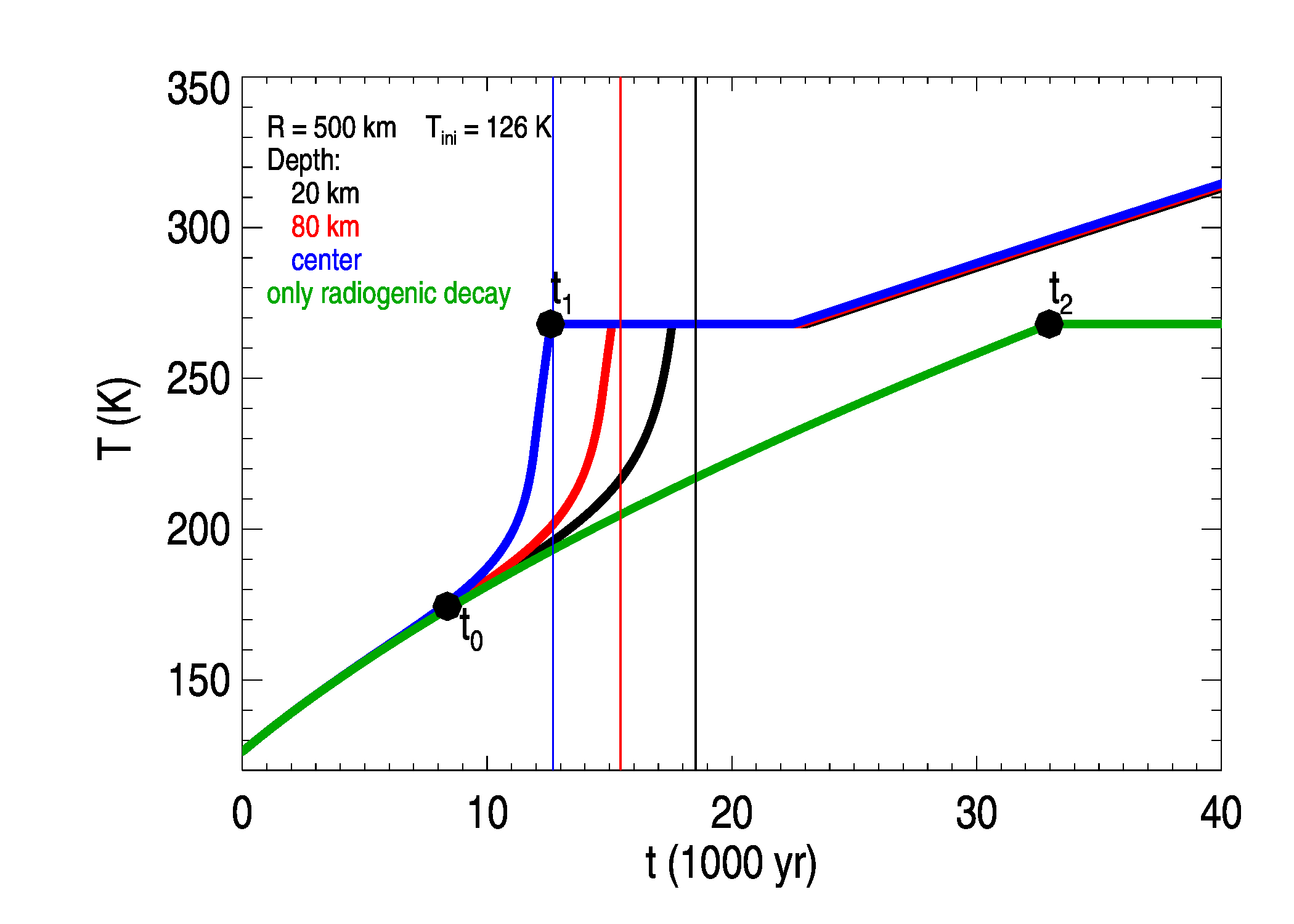}}}
\resizebox{7.8cm}{!}{\rotatebox{0}{\includegraphics{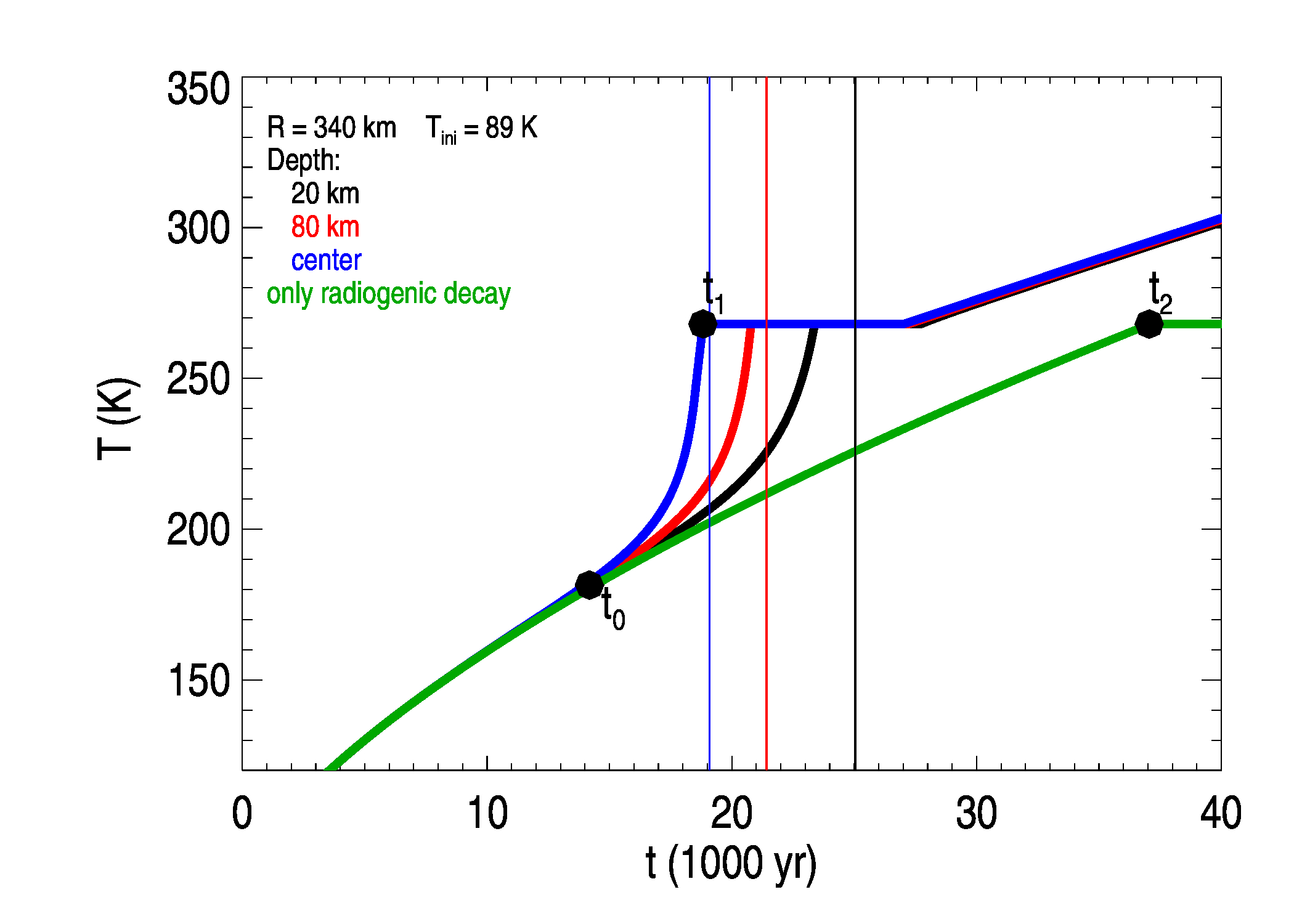}}}
}
\hbox{
\resizebox{7.8cm}{!}{\rotatebox{0}{\includegraphics{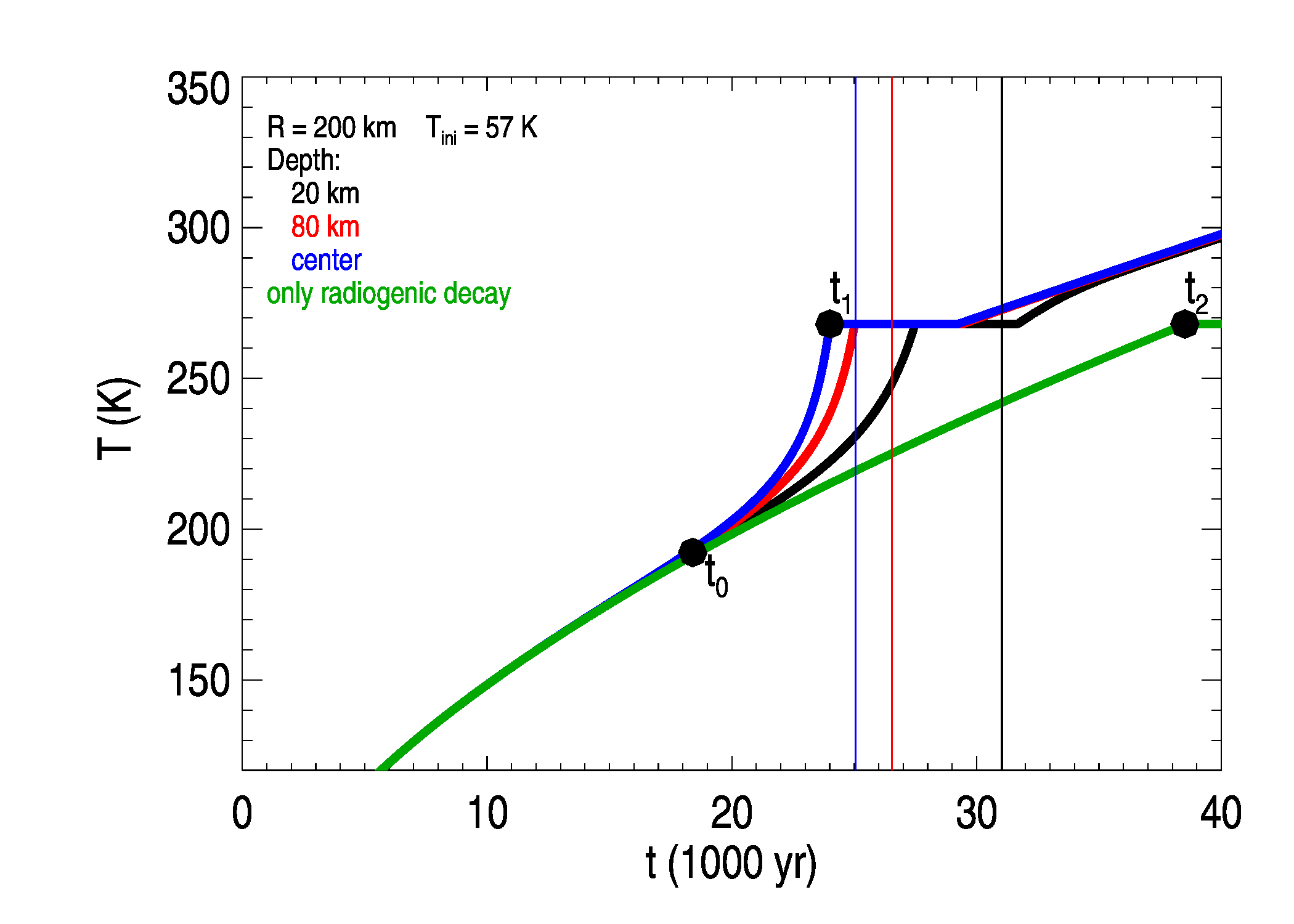}}}
\resizebox{7.8cm}{!}{\rotatebox{0}{\includegraphics{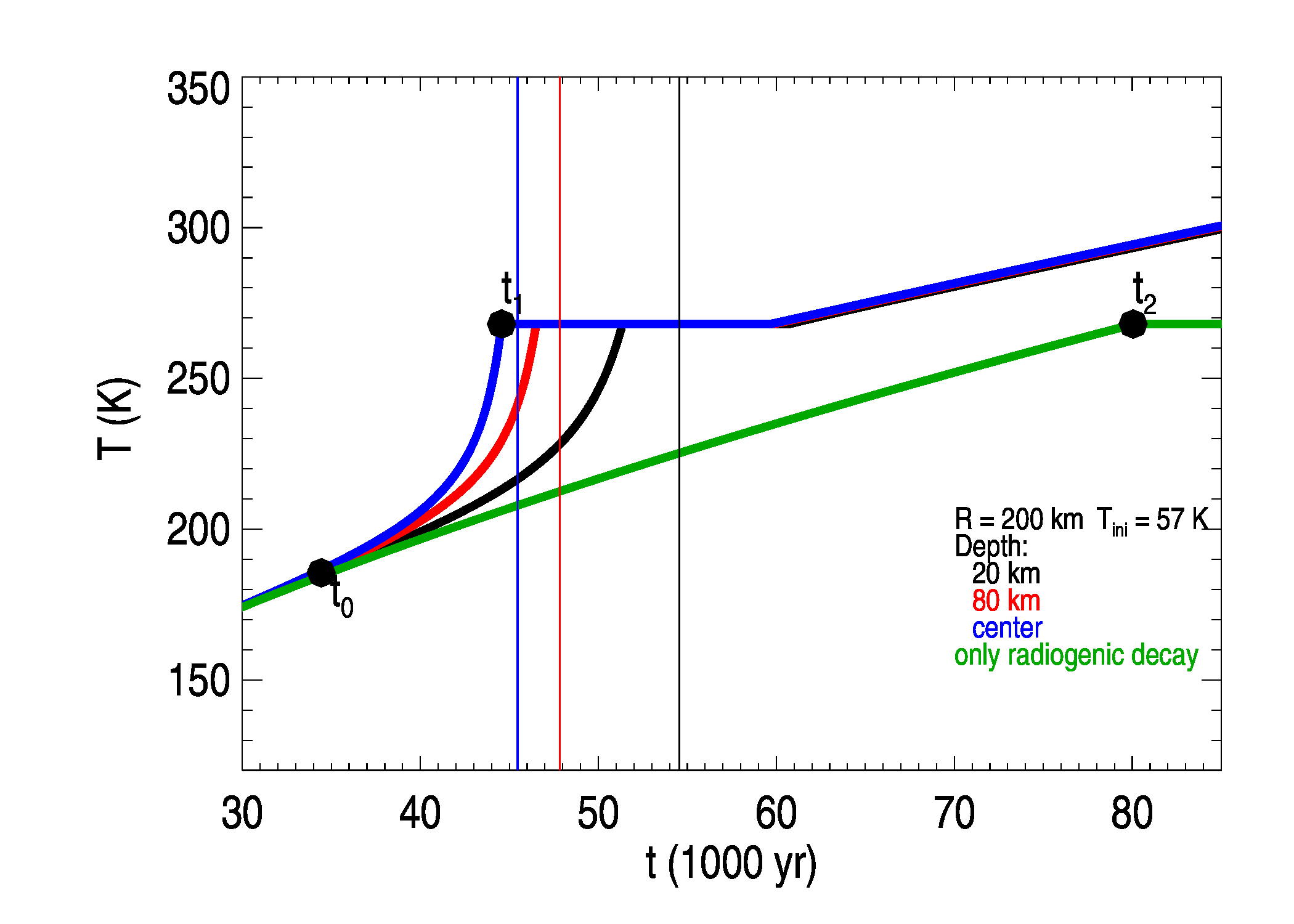}}}
}
\caption{Temperature evolution inside planetesimals with radii R\,=\,500, 340 and 200\,km (top-left, top-right and bottom-left, respectively), at different depth in their interior (20 and 80\,km, and at their core, as indicated in the figures). The green curves correspond to a thermal evolution {\it without} serpentinization at the core of the objects (this is representative for most layers due to the slow heat transport). In each subfigure three points are marked: 
t$_0$: the start of the serpentinization process; 
t$_1$: the time when that temperature reaches the melting point of water ice in the core, considering serpentinization; 
t$_2$: the time when that temperature reaches the melting point of water ice without serpentinization. 
The vertical lines represents the end of the chemical reaction.
In the bottom-right subfigure we present the results for an R\,=\,200\,km object, but assuming that the evolution starts 0.7\,Myr later, when the decay of $^{26}$Al produces only half of the heat compared with the object in the bottom-left subfigure. 
}

\label{fig:TNO_DT01}
\end{center}
\end{figure*}

\begin{figure}[!ht]
\begin{center}
\resizebox{7.8cm}{!}{\rotatebox{0}{\includegraphics{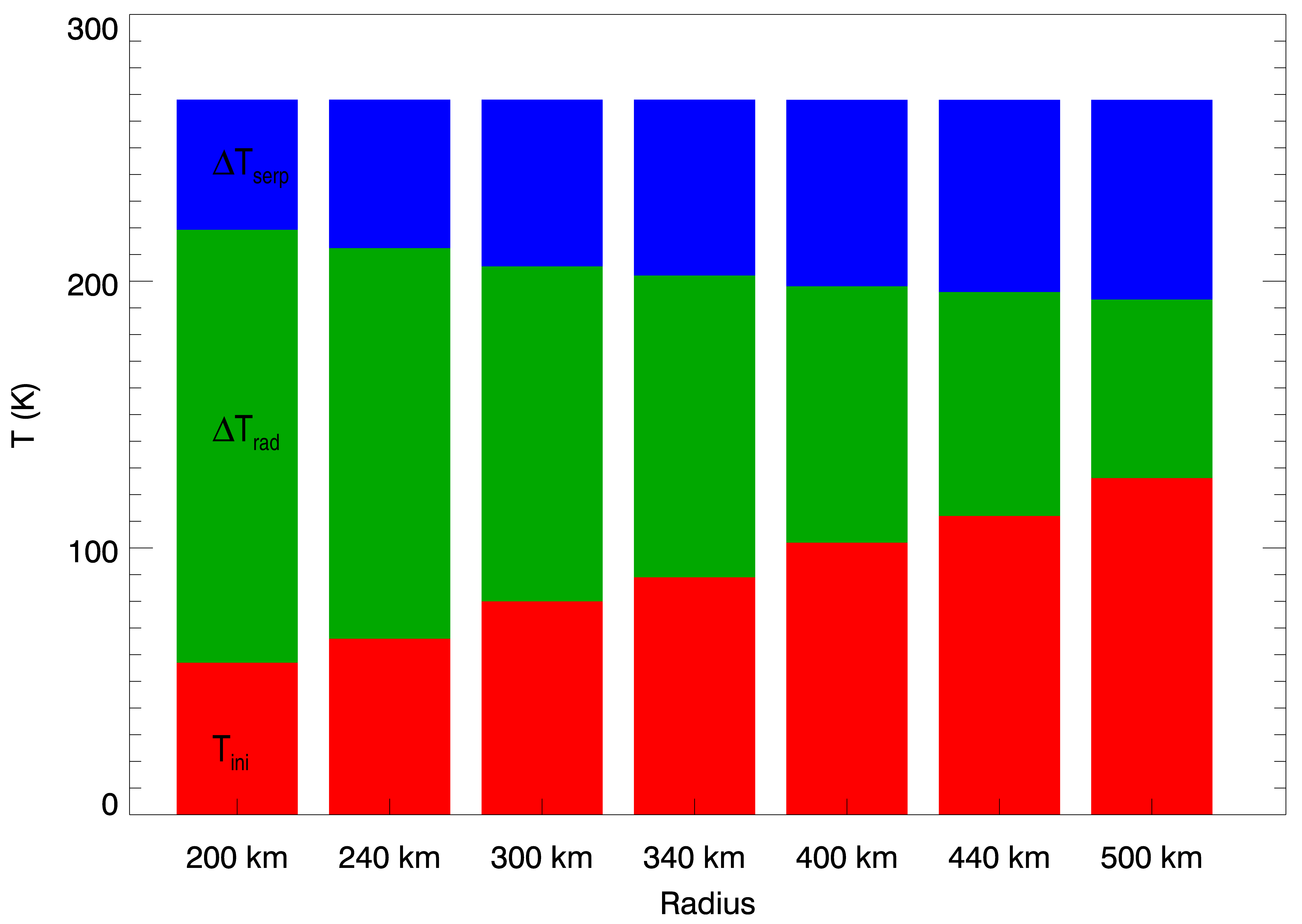}}}
\resizebox{7.8cm}{!}{\rotatebox{0}{\includegraphics{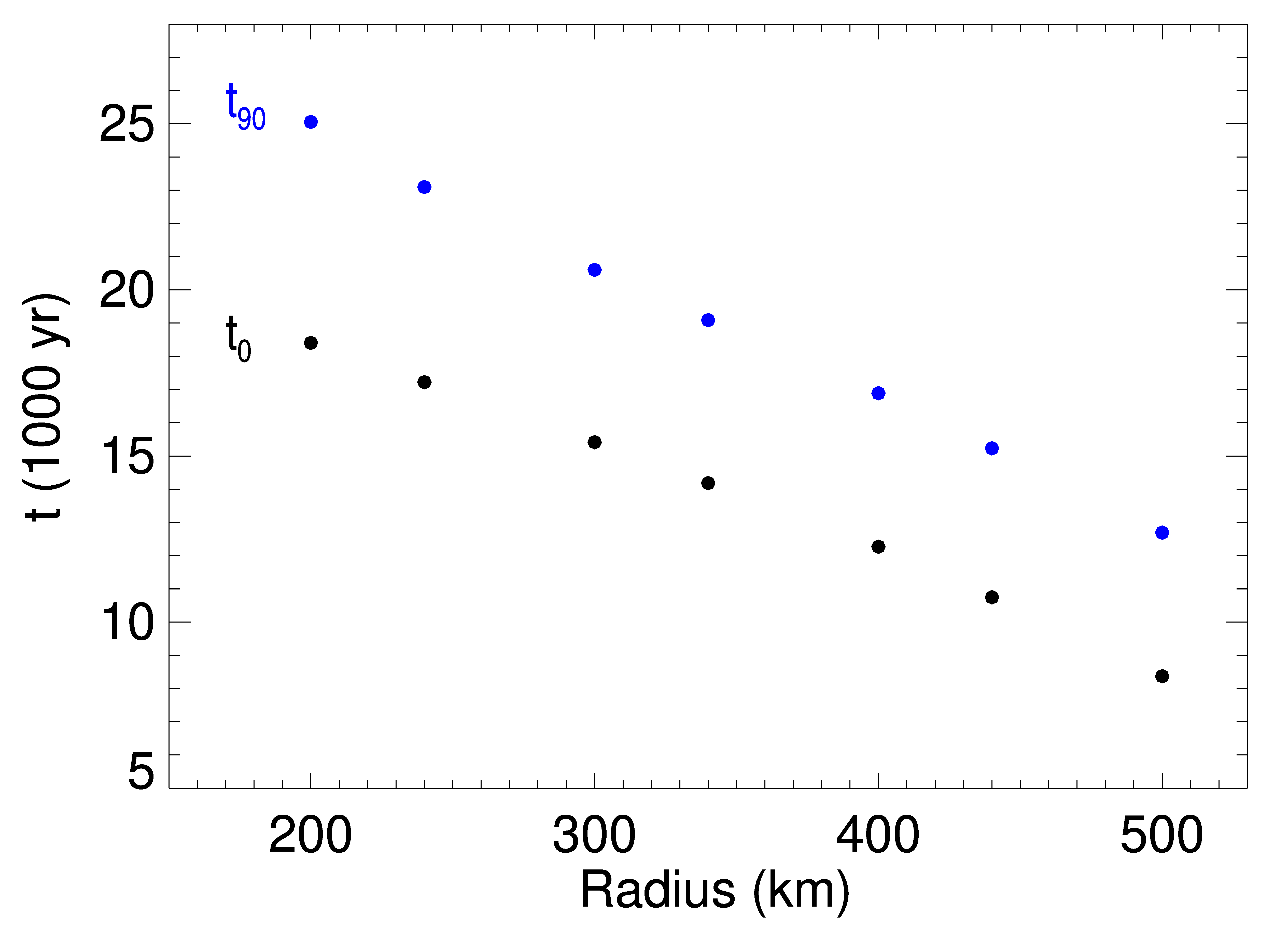}}}
\resizebox{7.8cm}{!}{\rotatebox{0}{\includegraphics{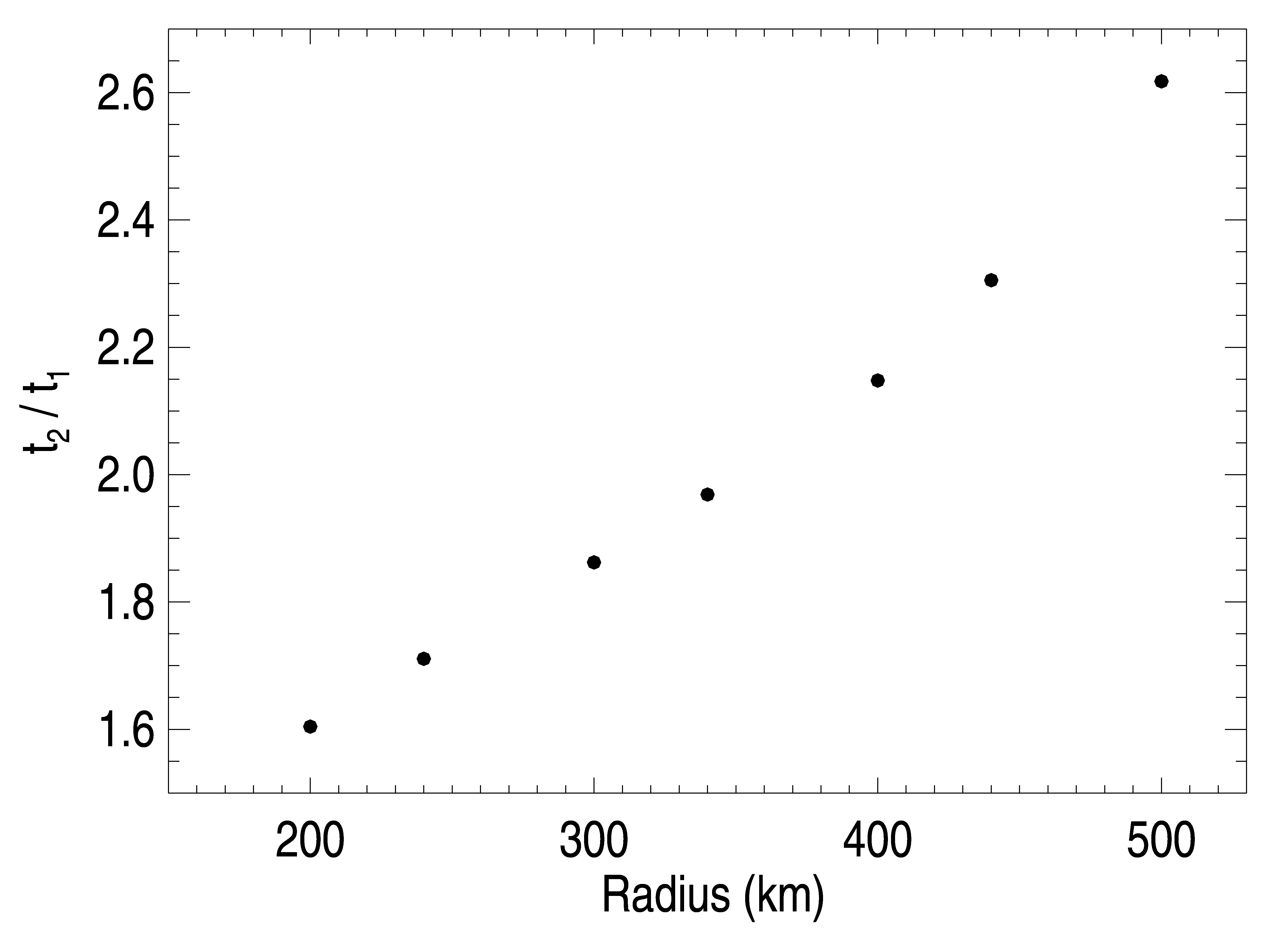}}}
\caption{Upper panel: Contribution of accretion heat (red), radiogenic decay (green) and serpentinization (blue) to the final temperature of the test object at the time when the melting temperature of the water ice is reached in the center;
Middle panel: t$_{90}$ and t$_0$ versus object radius; 
Bottom panel: The ratio of the times necessary to reach the melting temperature of water ice, without and with serpentinization (t$_2$/t$_1$), as a function of the radius of the object.
}
\label{fig:TNO_DT02}
\end{center}
\end{figure}

To investigate this, we modeled the thermal evolution of test objects with  radii between 150\,$<$\,R\,$<$\,500\,km considering all heat sources (accretion heat, radiogenic decay) in addition to serpentinization itself. In Fig.~\ref{fig:TNO_DT01} we present the thermal evolution of three objects with radii 200, 340 and 500\,km. The serpentinization process starts to produce a significant amount of heat when the temperature of the layer exceeds a critical value of $\sim$180\,K, at time t$_0$ after the start of the simulation. This t$_0$ depends strongly on the size of the object (see Fig.~\ref{fig:TNO_DT02}, middle panel) and decreases with size, in a similar way as the end time of the reaction (t$_{90}$), represented by vertical lines in Fig.~\ref{fig:TNO_DT01}. 
In these simulations serpentinization completes fully 
in all the studied layers even at the melting point, and therefore it does not contribute to the future thermal evolution. We note that all our simulations have been run beyond the time of completion of the reaction. In our simulations serpentinization reduces the time needed to reach the melting temperature of water by a factor of 1.6-2.6, depending on the size of the object (Fig.~\ref{fig:TNO_DT02}, bottom panel). The contribution of the different processes to the temperature of the object at the end of serpentinization (t$_{90}$) is presented in Fig.~\ref{fig:TNO_DT02} (upper panel). 
At smaller sizes radiogenic decay contributes the most, and the importance of accretion heat increases notably with size. The heat obtained from serpentinization (50-80\,K) becomes more important with growing size, and exceeds the contribution from radiogenic decay for R\,$\approx$\,500\,km. In large enough objects (where the critical reaction-starting temperature of $\sim$180\,K is reached) serpentinization proceeds quickly, and the whole process in finished in $\sim$10$^4$\,yr.  

The results of \cite{Wakita2011} shows that the formation time is very important because 2.4\,Myr after Ca-Al-rich inclusions formation, the icy planetesimals may not reach the melting temperature of ice. We also examined the effect of formation time on the thermal evolution, considering the same setup as above. 

In Fig.~\ref{fig:TNO_DT01}, bottom-right panel we demonstrate the effect of a delayed start (by 0.7\,Myr, the half-life of $^{26}$Al) and hence reduced radiogenic heat. In this simulation the 
$t_0$, $t_1$ and $t_2$ timescales are notably, by a factor of $\sim$2 longer.

Despite these longer timescales, serpentinization can still fully proceed, and the temperature reaches the melting point of ice. As it is expected, the contribution from radiogenic decay to the final temperature is smaller (11\,K in this specific case) and the relative contribution of the temperature increase due to serpentinization is higher.  

\begin{figure}[!ht]
\begin{center}
\resizebox{8cm}{!}{\rotatebox{00}{
\includegraphics{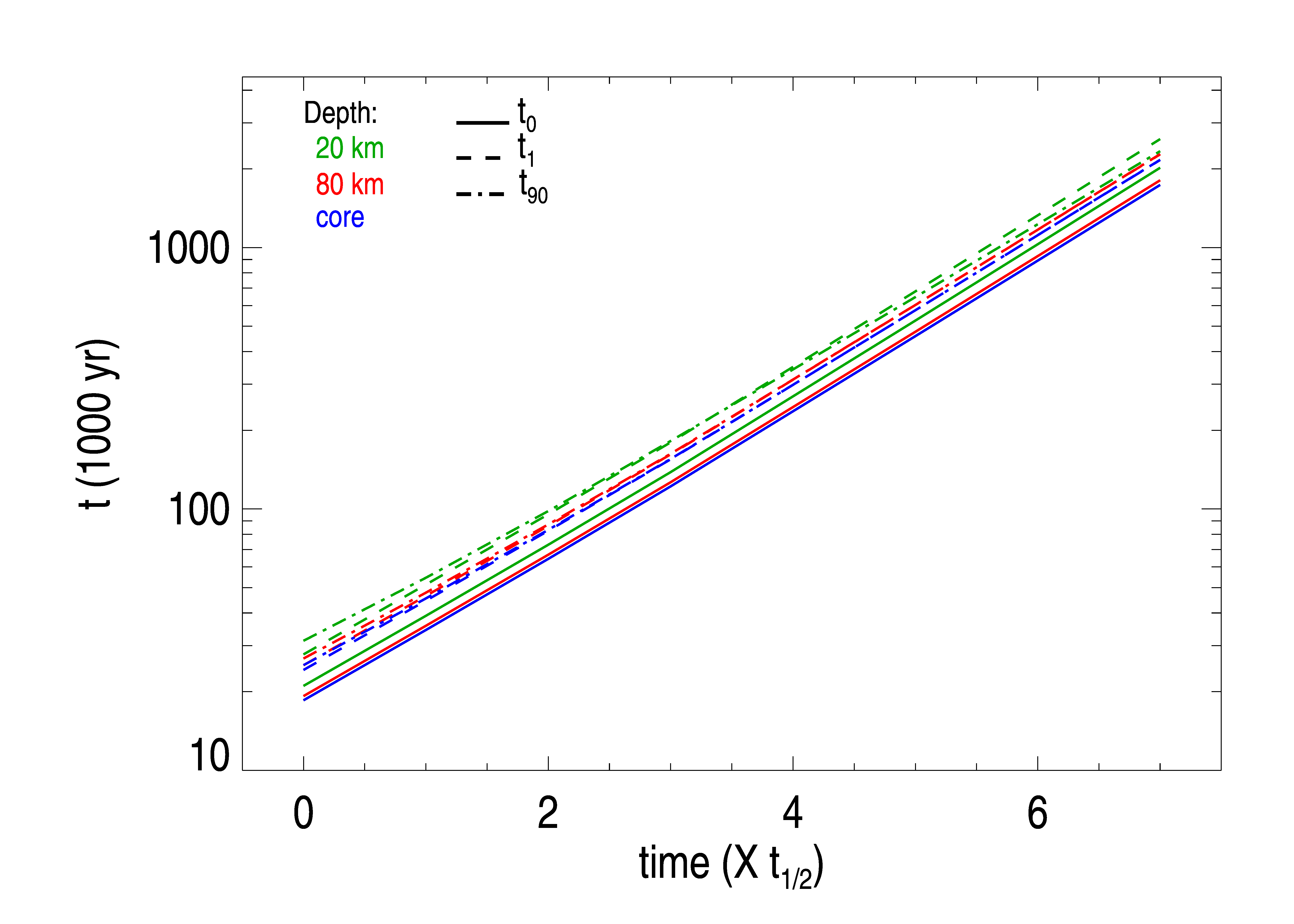}
}}
\caption{t$_0$ (solid curve), t$_1$ (dashed) and t$_{90}$ (dash-dotted) times of the simulations of the thermal evolution of an R\,=\,200\,km object versus the starting time relative to the original starting time when $^{26}$Al abundance was at its maximum. Formation time (the start time of the simulation) is presented as multiples of the half-life of $^{26}$Al (t$_{1/2}$). The last data point at 7$\times$0.717\,Myr\,$\approx$\,5\,Myr corresponds to the expected maximum lifetime of the protoplanetary disk. Green, red and blue colours correspond to 20 and 80\,km depths, and the core of the planetesimal, respectively.}
\label{fig:TNO_halflife}
\end{center}
\end{figure}
In Fig.~\ref{fig:TNO_halflife} we show the effect of a late formation of the planetesimals, and consequently a late start of the serpentinization process, for different starting times. The late formation results in a reduced amount of heat from radiogenic decay, due to the depletion of $^{26}$Al. These simulations were performed for R\,=\,200\,km-sized objects, assuming that the formation of the planetesimal happened at a t\,=\,[0,1,...7]$\times$t$_{1/2}$ after the onset of isotopic decay, when the abundance of $^{26}$Al was at its maximum, as considered in our previous simulations.
Serpentinization needs a much longer time to start, either in the core or in layers closer to the surface, for later formation times. While this t$_0$ is in the order of a few thousand years for an early formation, it is several million years for a formation at t\,$\approx$5\,Myr. 
The t$_{90}$ timescale, i.e. the time when 90\% of the potentially serpentine forming material in the core of the planetesimal is consumed, is also notably longer, while the {\it ratio} of t$_0$ to t$_{90}$ remains roughly the same. Altogether the lower temperatures due to the smaller radiogenic heat slow down serpentinization considerably.



\begin{figure}[!ht]
\begin{center}
\resizebox{7.8cm}{!}{\rotatebox{0}{\includegraphics{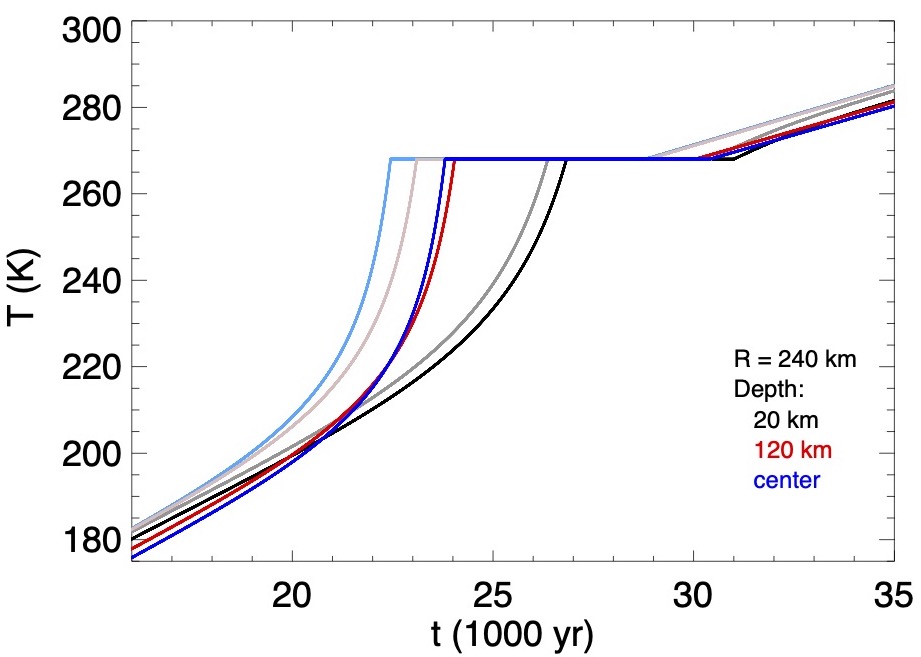}}}
\caption{Thermal evolution of a planetesimal of R\,=\,240\,km with an outward temperature gradient at the start, represented by curves with 'normal' colours.  
The 'pale' colours correspond to the same object/layer, but assuming a homogeneous temperature distribution at the start of the calculations.}
\label{fig:inhom}
\end{center}
\end{figure}

We examined what changes occur when the initial temperature distribution is inhomogeneous and the surface is warmer (Fig.~\ref{fig:inhom}). In this test we used an object with a radius of 240\,km and the initial temperature distribution was set in a way that the surface temperature was 5/3 times the central temperature \citep{Hanks1969}. Despite the lower starting temperature the serpentinization process is the fastest in the core due to the higher lithospheric pressure. The reaction takes longer in the outer layers, and as the initial temperature is higher, the final temperature will also be higher, but the temperature difference is smaller than in the homogeneous case.

\begin{figure*}[!ht]
\begin{center}
\hbox{
\resizebox{8cm}{!}{\rotatebox{0}{\includegraphics{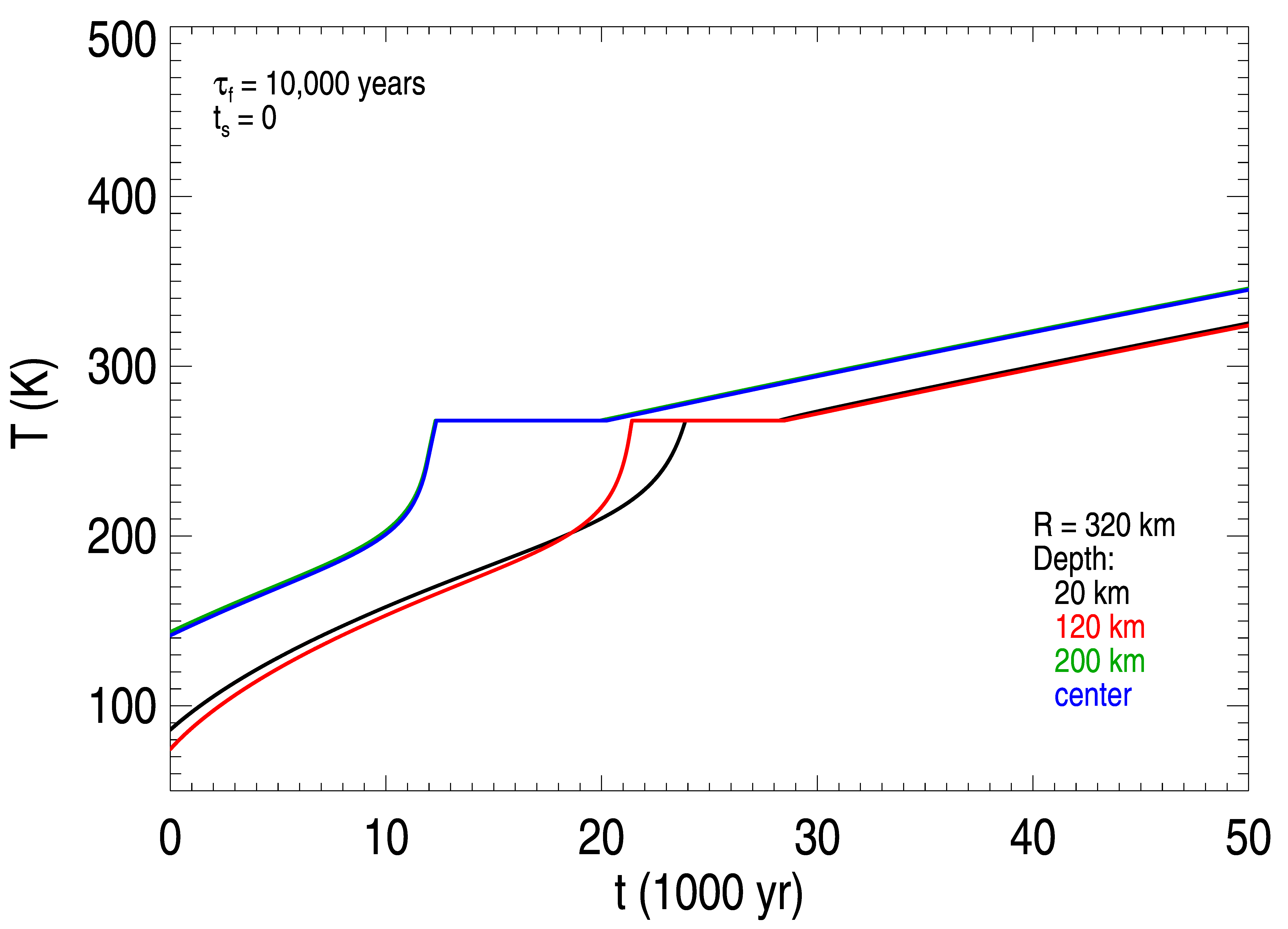}}}
\resizebox{8cm}{!}{\rotatebox{0}{\includegraphics{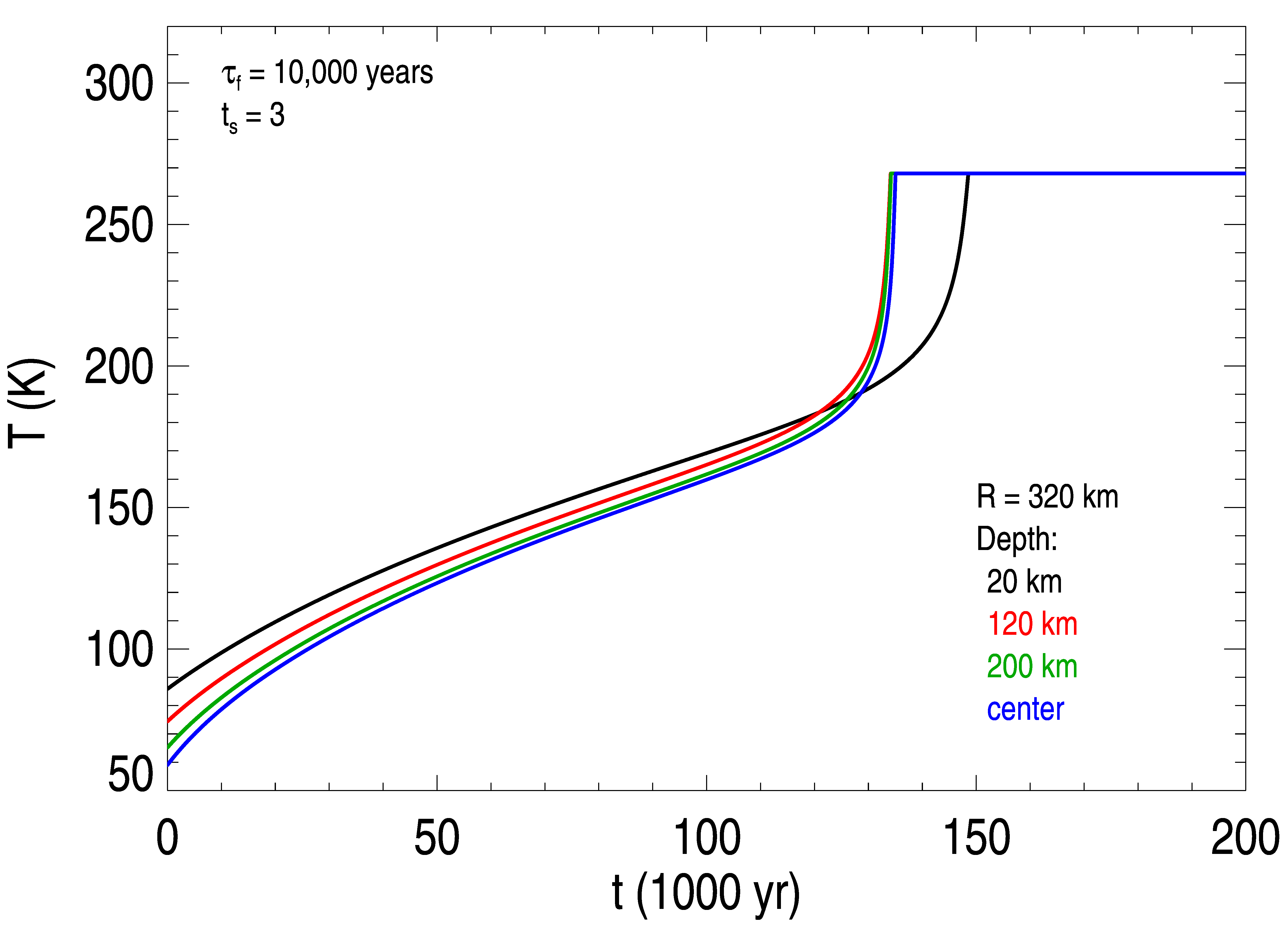}}}
}
\hbox{
\resizebox{8cm}{!}{\rotatebox{0}{\includegraphics{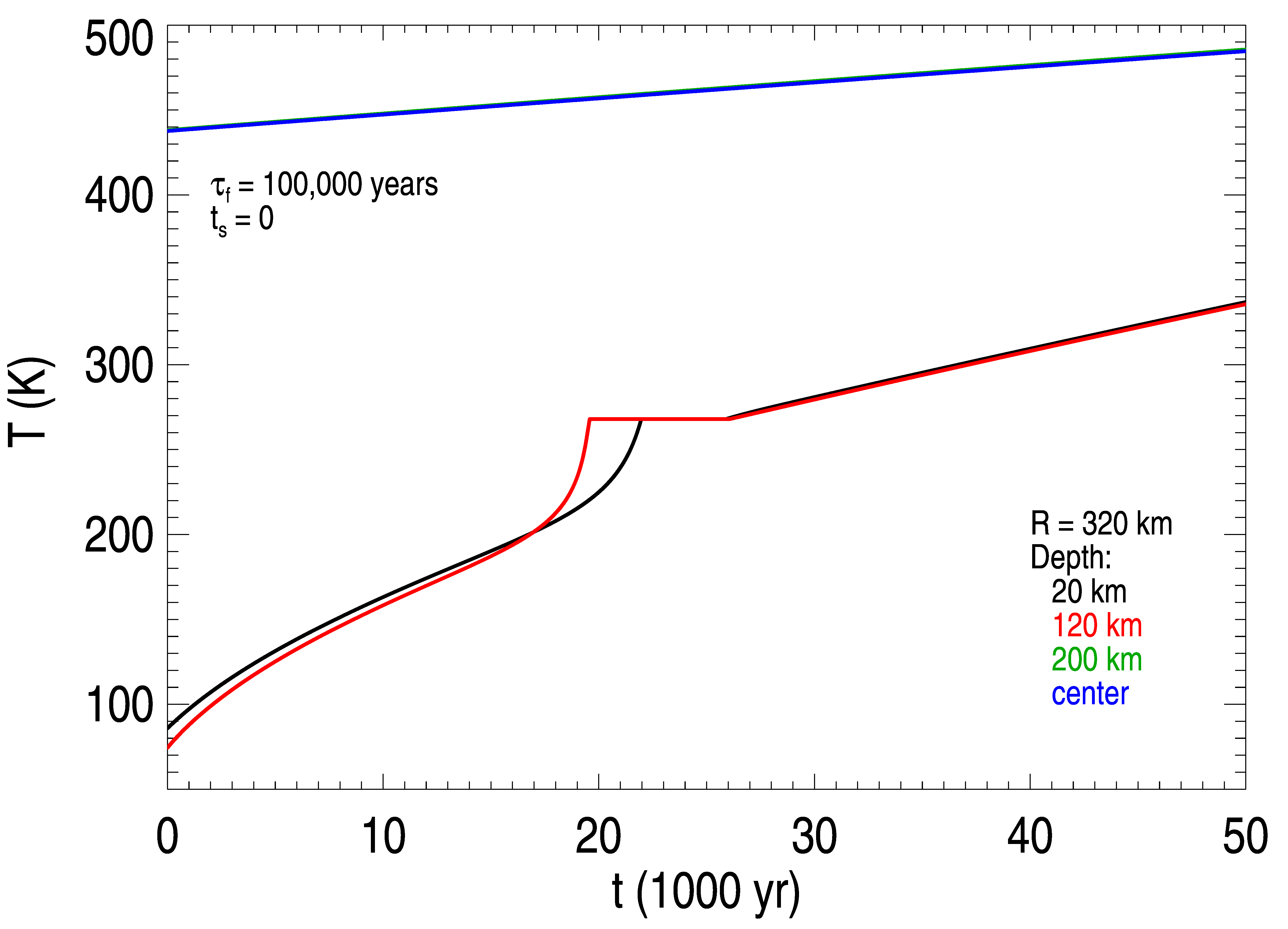}}}
\resizebox{8cm}{!}{\rotatebox{0}{\includegraphics{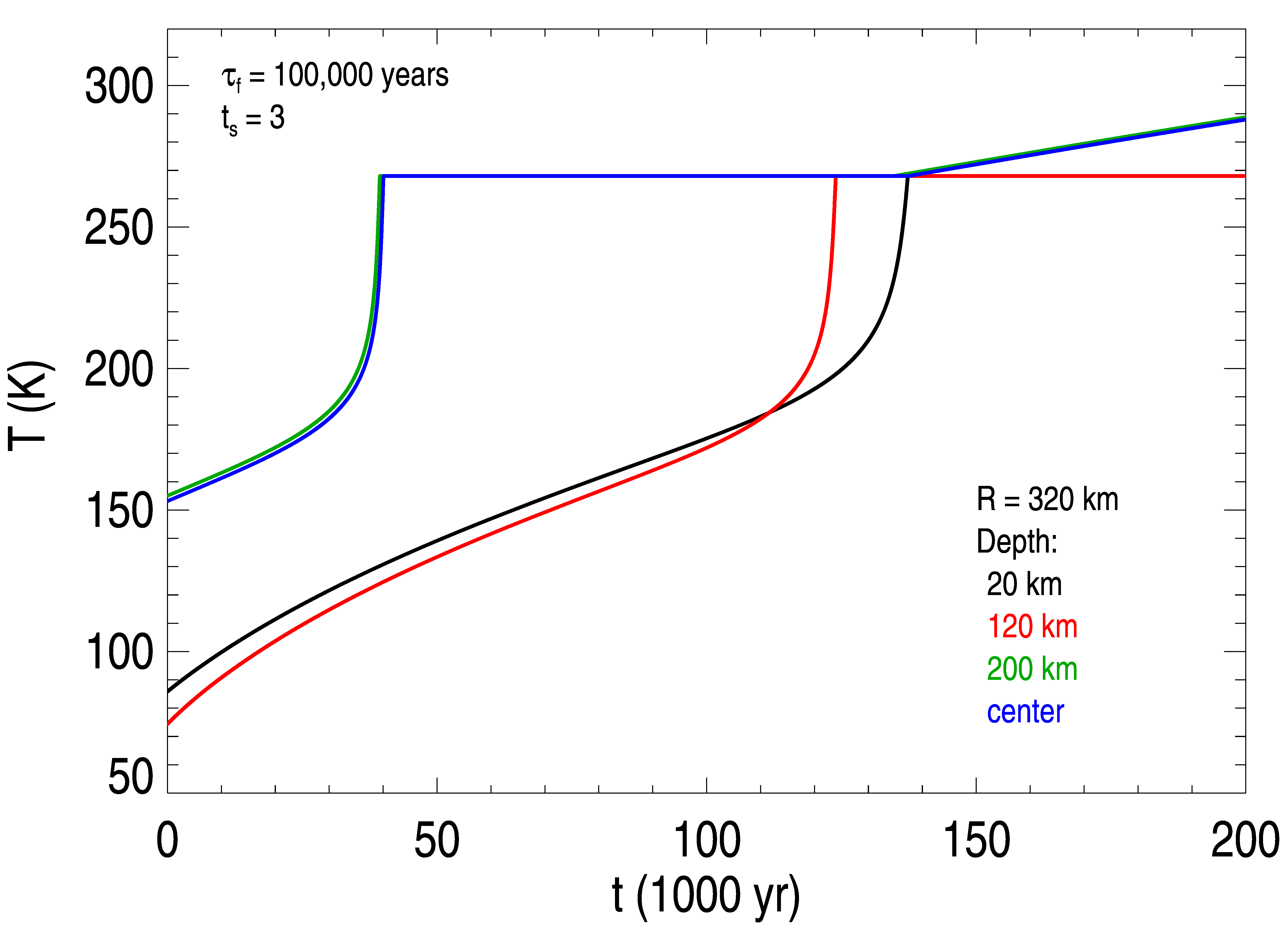}}}
}
\caption{In this figure we present the temperature evolution obtained in models with a two-phase formation scenario. In the top row the first formation/accretion event is followed by another event in $\tau_{f}$\,=\,10,000\,yr, while in the bottom row in  $\tau_{f}$\,=\,100,000\,yr. On the left the first formation event occurs early, at the maximum of the radiogenic heat production of $^{26}$Al; on the right the first formation event occurs three $^{26}$Al half-life later. t\,=\,0 corresponds to the second formation event.}
\label{fig:inhom_2step}
\end{center}
\end{figure*}

We also examined the case when the initial temperature distribution was inhomogeneous and the formation occured in two steps (Fig.~\ref{fig:inhom_2step}). First, an object formed with a radius of 160\,km, and $\tau_f$\,=\,10,000 or 100,000\,yr later the formation was completed and the object reached a final radius of 320\,km. In the time between the two formation/accretion events the object may have warmed up both from radiogenic decay and serpentinization. We assumed two values for the time of the first formation event: t$_s$\,=\,0, the maximum $^{26}$Al heat production date, and t$_s$\,=\,3\,t$_{1/2}$, i.e. three $^{26}$Al half-life later. 
For $\tau_s$\,=\,100,000\,yr and t$_s$\,=\,0 the serpentinization could go along all the way before the second accretion event in the centre and left the core at a high temperature. Due to the lack of further reactants, the core was heated by radiogenic decay only from this point on (almost straight blue curve in Fig.~\ref{fig:inhom_2step}, bottom left). The final object is built on this warm core in the second accretion event. 
Also in our other three cases, the results show that serpentinization is faster in the core independently of the initial conditions. The difference between the core and the outer layers is more significant for t$_s$\,=\,0, and, as expected, everything occurs significantly later for t$_{s}$\,=\,3\,t$_{1/2}$\,yr. 

\section{Conclusions}

We presented an improved algorithm to model the serpentinization reaction and its role in the thermal evolution of planetesimals in the early Solar system, based on the model by \cite{Gobi2017}. In our model we incorporated several previously overlooked or neglected effects: (i) the calculation of heat capacity of water is more accurate in this work; (ii) we have taken into account the depth dependence of lithospheric pressure; (iii) the latent heat of vaporization of water was calculated in a more exact way, and (iv) we improved the method to calculate the amount of microscopic liquid interfacial water at subzero temperatures which eventually allowed a smaller amount of water to react compared with previous model. 
Our model is able to follow the local chemical evolution of the serpentinization process and the algorithm was inserted into a more complex internal heat evolution model which considered radioactive decay and heat transfer. We demonstrated that the improvements in our models lead to results different from those of earlier models, both in serpentinization reaction timescales and in final heat production. 
The presence of interfacial water at temperatures below the melting point of bulk water ice may be able to start the reaction, though at a lower rate, and may eventually be able to produce a notable temperature increase. 

Our results suggest that there is a size limit of R\,$\approx$\,600\,km (assuming Kuiper belt compositions) above which the serpentinization process becomes efficient, mainly due to the higher accretion heat of these larger objects. This is even true in later times when radiogenic decay cannot significantly contribute to the thermal budget because of the depletion of $^{26}$Al, several million years after the onset of radiogenic decay. Serpentinization may proceed in smaller objects after the onset of radiogenic decay (down to $\sim$150\,km) in the presence of notable heat from radiogenic decay. The overall importance of serpentinization in the chemical evolution of smaller (R\,$\leq$\,500\,km) objects in the outer Solar system depends strongly on the time the process starts at, as accretion heat alone cannot start the process for these smaller planetesimals. The lifetime of the planetesimal forming disks around solar-like young stars is $\sim$3-5\,Myr, and very few disks survive 10\,Myr \citep[see e.g.][for a summary]{Russell}. As we showed above, the reaction timescales and the overall efficacy depend strongly on the heat provided by radiogenic decay, and it is reduced strongly if planetesimal formation is delayed. However, serpentinization is still able to proceed, although with a notably reduced speed, even if the object is formed a few million years later, or in multiple accretion events, at the smallest sizes we investigated (R\,=\,200\,km). 

The bulk serpentinization efficiency -- the ratio of objects in which serpentinization reformed the interior and those where it could potentially do it so -- also depends strongly on the collisional evolution of planetesimals in the young trans-Neptunian region. While serpentization is a fast process in the case of an early formation, it can be considerably slower if the formation process is delayed. A simple estimate based on \citet{Wyatt} shows that for objects with R\,$\ge$\,100\,km destructive collisions occur on the million year timescales for a wide range of possible disk parameters (disk mass, disk extension, mean orbital eccentricity, material strength, etc.), and this timescale gets longer with the disk dispersal. In this sense the serpentinization timescale is expected to be much shorter than the collisional timescale at any time while the disk exists. 
In our Solar system the trans-Neptunian population of objects with radii R\,$>$\,30\,km are expected to be primordial, as they 'decoupled' early from the collisional evolution, at the end of the runaway growth, and before the onset of destructive collisions of smaller objects \citep{Schlichting2013}. This suggests that our R\,=\,150-500\,km objects, once formed, are likely not destroyed by collisions, and serpentinization could take place in their interior.  

%

\section*{Acknowledgements}

This research has been supported by the K-125015, K-138962 and K-138594 grants of the Hungarian Research, Development, and Innovation Office (NKFIH). We are indebted to our reviewers for their comments and suggestions which have helped us to notably improve this paper. 

\appendix
\section{Serpentinization model}\label{app:serp}

In this model a specific object was considered to be made of seven separate components: olivine, enstatite, non-reactive solid, serpentinite, and water in three-phase state as liquid, solid, and vapor in the pore space.
These materials are marked in the subscripts as: $oli$, $ens$, $nre$, $ser$, $wat$, $ice$ and $vap$.
Table\,\ref{tab:in_val} presents the variables and constants used in the serpentinization model.
In those cases when the initial values weren't specified otherwise, we used the initial values from Table\,\ref{tab:in_val}.

\begin{table*}[ht!]
    \centering
    \caption{List of variables and constants 1.}
    \begin{tabular}{ll}
    \\
    \hline
$C_p$ & Heat capacity ($Jkg^{-1}K^{-1}$) \\
$g$ & Gravitational acceleration ($m/s^2$) \\
$\textit{k}_r$ & Observable serpentinization rate (mol/year)\\
m & Mass of components ($kg$) \\
n & Amount of substance (mol) \\
$\rho$ & Density ($kg/m^3$) \\
$\phi$ & Porosity \\
$r_{lt}$ & Top of the layer ($m$)\\
$r_{lb}$ & Bottom of the layer ($m$)\\
$P_{lit}$ & Lithostatic pressure (Pa) \\
$P_{vap}$ & Pressure of water vapor (Pa) \\
$T$ & Temperature ($K$) \\
$t$ & Time ($year$) \\
V & Volume of components ($m^3$) \\
$\Delta H_{v}$ & Latent heat of vaporization of water ($J/kg$)\\
$\Delta H_{r}$ & reaction enthalpy ($J/mol$) \\
$\Delta h_{vap}$ & vaporization heat ($J$) \\
$\Delta h_{ser}$ & reaction heat ($J$) \\
$\Delta n_{ser}$ & Serpentine produced ($mol$)\\
$\Delta T_{ser}$ & Temperature rise from the serpentinization \\
$w$ & Content of microscopic liquid water ($g/100\,g soil$) \\
\hline
$\lambda$ & decay constant ($sec^{-1}$) \\
$q_{rad}$ & radiogenic heat production rate ($W/kg$) \\
$Q_{rad}$ & radiogenic heating rate ($J$) \\
$k$ & thermal konductivity ($W/m/K$) \\
    \hline
    \end{tabular}
    \label{tab:in_val}
\end{table*}

\begin{table*}[ht!]
    \centering
    \caption{List of variables and constants 2.}
    \begin{tabular}{ll|ll}
    \\
\hline
&& constant & initial values \\
\hline
\hline
$G$ & Gravitational constant ($m^3 kg^{-1} s^{-2}$) & $6.67408\times10^{-11}$ \\
$\varepsilon$ & Emissivity & 0.9 &\\
$\sigma$ & Stefan-Boltzmann Constant ($W/m^2/K^4$) & $5.6697\times10^{-8}$ \\
$L$ & Latent heat of fusion of ice ($J kg^{-1}$) & $3.3\times10^5$ & \\
$R_g$ &  Gas constant ($Jmol^{-1}K^{-1}$) & $8.314$ \\
$\rho_{oli}$ & Density of olivine ($kg m^{-3}$) & 3210 \\
$\rho_{ens}$ & Density of enstatite ($kg m^{-3}$) & 3190 \\
$\rho_{nre}$ & Density of non-reactive solid ($kg m^{-3}$) & 3630 \\
$\rho_{ser}$ & Density of serpentinite ($kg m^{-3}$) & 2470 \\
$S$ & Specific surface area ($m^2g^{-1}$) & 100 & \\
$T_{melt}$ & Melting point of water ($K$) & 268\,K\\
\hline
$t_{1/2}$ & Half-life of $^{26}Al$ ($Myr$) & 0.717 & \\
$TE$ & Total energy of $^{26}Al$ ($J/kg$) & $5.07 \times 10^6$ & \\
$k_{oli}$ & thermal conductivity of olivine ($W/m/K$) & 5.155 & \\
$k_{ens}$ & thermal conductivity of enstatite ($W/m/K$) & 5.155 & \\
$k_{ser}$ & thermal conductivity of serpentinite ($W/m/K$) & 2.95 & \\
$k_{nre}$ & thermal conductivity of non-reactive solid ($W/m/K$) & 2.8 & \\
$T_{amb}$ & Ambient temperature ($K$) & 50 & \\
\hline
\hline
$n_{oli}/n_{wat}$ & Olivine to water ratio && 1:2, 0.12 \\
$n_{nre}$ & Non-reactive material ($\%$) && 14 \\
$R$ & Radius of planetesimal  ($km$) & & 15 - 800 \\
$\Delta t$ & Time step ($year$) && 0.5\\
\hline
    \hline
    \end{tabular}
    \label{tab:in_val}
\end{table*}
\subsection{Material properties: heat capacity and density of components}\label{app_prop}

Heat capacity ($c_p$) and density ($\rho$) calculations were obtained from \cite{Cohen2000}.
These physical properties of all components are temperature ($T$) dependent.
\begin{gather}
log\,C_p\,(olivine) = -11.32 + 13.58\,x - 4.25\,x^2 + 0.44\,x^3 \\
log\,C_p\,(enstatite) = -8.620 + 10.39\,x - 3.00\,x^2 + 0.28\,x^3 \\	
log\,C_p\,(water) = 8.250 - 4.180\,x + 1.12\,x^2 - 0.076\,x^3
\end{gather}
\noindent where $x=log\,T$. These expressions are valid from approximately 50 to 500\,K.
\begin{equation} 
\begin{split}
C_p\,(serpentinite) & = 1145+0.048\,T-2.65\times 10^7\,T^{-2 } \quad (T\,>\,273\,K) \\
log\,C_p\,(serpentinite)& = -0.59 - 1.51\,x + (2.82\,x^2) - (0.66\,x^3) \\
                    &  (T\,<\,273\,K)
\end{split}
\end{equation}
\begin{equation}
C_p\,(vapor) = 1730.54 + 0.45\,T \\
\end{equation}
\begin{equation} 
\begin{split}
C_p\,(ice) & = 152.46 + 7.12\,T \quad (T\,>\,150\,K) \\
 & = 126.89 + 7.50\,T \quad (150\,K\,>\,T\,>\,95\,K) \\
 & = -49.97 + 9.5\,T \quad (95\,K\,>\,T\,>\,50\,K)
\end{split}
\end{equation}
For a specific layer of the object, we calculated the average heat capacity from the components' heat capacities and masses ($m$):
\begin{equation} 
\begin{split}
C_p\,(layer) m_{lay} & = C_p\,(oli)\,m_{oli} + C_p\,(ens)\,m_{ens} + C_p\,(ice)\,m_{ice} \\
        &  + C_p\,(wat)\,m_{wat} + C_p\,(vap)\,m_{vap}  \\
        &  + C_p\,(ser)\,m_{ser} + C_p\,(oli)\,m_{nre}
\end{split}
\end{equation}

In the cases of olivine, enstatite, serpentinite and non-reactive material, constant values of density were used (see in Table\,\ref{tab:in_val}).
For \ce{H2O} we used:
\begin{equation}
\rho_{wat} = -221 + 13.1\,T - 0.0507\,T^2 + 8.49\times10^{-5}\,T^3 - 5.48\times10^{-8}\,T^4
\end{equation}
\begin{equation}
\begin{split}
\rho_{ice} & = -46.9\,x+1032.71 \quad (T\,>\,137\,K) \\
  & = -1.32\,x+935.32 \quad (T\,<\,137\,K)
\end{split}
\end{equation}
\begin{equation}
\rho_{vap} = P_{vap} \, 0.018 \, T^{-1} \, R_g^{-1}
\end{equation}
\noindent where $P_{vap}$ is the vapor pressure of the vapour and $R_g$ is the gas constant. For the examined layer, we calculated the average density from the components' masses and volumes ($V$):
\begin{equation}
\rho_{lay} = \frac{m_{oli} + m_{ens} + m_{ice} + m_{wat} + m_{vap} + m_{ser} + m_{nre}}{V_{oli} + V_{ens} + V_{ice} + V_{wat} + V_{ser} + V_{voi} + V_{nre}}
\end{equation}
\noindent where $V_{voi}$ is the void space  which is filled with \ce{H2O} vapor.
\subsection{Mass and volume of components}\label{app_mass}

 This part of the algorithm considers a specific layer in a body, at a specific depth. 
The volume of the examined layer is as follows:
\begin{equation}
V_{lay} = \bigg(\frac{4}{3} \, \pi\,r_{lt}^3 \bigg)-\bigg(\frac{4}{3} \, \pi\,r_{lb}^3\bigg)
\end{equation}
\noindent where $V_{lay}$ is the volume of the examined layer, $r_{lt}$ is the top of the layer and $r_{lb}$ is the bottom of the layer.

The initial values of volumes and masses of components (marked with $(0)$) were calculated from the porosity ($\phi$), olivine-to-water ratio ($n_{oli}/n_{wat}$) and the fraction of non-reactive material ($n_{nre}$):

\begin{gather}
V_{voi}(0) = V_{lay} \,\phi\\
V_{nre} = V_{lay} \, n_{nre} \\
m_{nre} = V_{nre} \, \rho_{nre} \\
m_{oli}/m_{wat}(0) = n_{oli}/n_{wat} \, 140.71 / 18 \\
m_{lay} = V_{lay} \, \rho_{lay}
\end{gather}
\begin{gather}
\begin{split}
V_{ens}(0) & = (V_{lay}-(V_{nre}+V_{voi})) \frac{m_{oli}/m_{wat}/1.3/\rho_{ens}}{\frac{m_{oli}/m_{wat}}{\rho_{oli}}+\frac{1}{\rho_{ice}}+\frac{m_{oli}/m_{wat}/1.3}{\rho_{ens}}} \quad (T\,<\,T_{melt}) \\
        & = (V_{lay}-(V_{nre}+{V_{voi}}) \frac{m_{oli}/m_{wat}/1.3/\rho_{ens}}{\frac{m_{oli}/m_{wat}}{\rho_{oli}}+\frac{1}{\rho_{wat}}+\frac{m_{oli}/m_{wat}/1.3}{\rho_{ens}}} \quad (T\,>\,T_{melt})
\end{split}
\end{gather}
\begin{gather}
\begin{split}
V_{ice}(0) & = (V_{lay}-(V_{nre}+V_{voi})) \frac{1/\rho_{ice}}{\frac{m_{oli}/m_{wat}}{\rho_{oli}} + \frac{1}{\rho_{ice}} + \frac{m_{oli}/m_{wat}/1.3}{\rho_{ens}}} \quad (T\,<\,T_{melt}) \\
            & = 0 \quad (T\,>\,T_{melt}) \\
\end{split}
\end{gather}
\begin{gather}
\begin{split}
V_{wat}(0) & = 0 \quad (T\,<\,T_{melt}) \\
        & = (V_{lay}-(V_{nre}+V_{voi})) \frac{1/\rho_{wat}}{\frac{m_{oli}/m_{wat}}{\rho_{oli}} + \frac{1}{\rho_{wat}} + \frac{m_{oli}/m_{wat}/1.3}{\rho_{ens}}} \quad (T\,>\,T_{melt})
\end{split}
\end{gather}
\begin{gather}
\begin{split}
V_{oli}(0) & = (V_{lay}-(V_{nre}+V_{voi})) \frac{m_{oli}/m_{wat}/\rho_{oli}}{\frac{m_{oli}/m_{wat}}{\rho_{oli}} + \frac{1}{\rho_{ice}} + \frac{m_{oli}/m_{wat}/1.3}{\rho_{ens}}} \quad (T\,<\,T_{melt}) \\
        & = (V_{lay}-(V_{nre}+V_{voi})) \frac{m_{oli}/m_{wat}/\rho_{oli}}{\frac{m_{oli}/m_{wat}}{\rho_{oli}} + \frac{1}{\rho_{wat}} + \frac{m_{oli}/m_{wat}/1.3}{\rho_{ens}}} \quad (T\,>\,T_{melt})
\end{split}
\end{gather}
The melting point of ice ($T_{melt}$\,=\,268\,K) is obtained considering a saturated solution of \ce{MgSO4}. 

The masses of components (enstatite, ice, water, olivine)  are calculated in the zeroth time step in the following way:
\begin{gather}
m_j(0) = V_j(0)\,\rho_j
\end{gather}
\noindent where $j$ subscript refers to the materials.
%
In all other time steps these masses are obtained as:
\begin{gather}
m_{ser} = m_{ser}^{i-1}+\frac{277.1\,\Delta n_{ser}^{i-1}}{1000} \\
m_{oli} = m_{oli}^{i-1} - (\Delta n_{ser}^{i-1}\times 0.14071) \\
m_{ens} = m_{ens}^{i-1} - (\Delta n_{ser}^{i-1}0\times 0.10039) \\
m_{ice} = max(0,m_{ice}^{i-1}-\Delta m_{ice}^{i-1}) \\
m_{wat} = m_{wat}^{i-1}+\Delta m_{ice}^{i-1} - 2\,\Delta n_{ser}^{i-1}\times 0.018) \\
m_{vap} = \frac{0.018 \, P_{vap} \, V_{voi}}{R_g \, T} \\
\end{gather}

 where the $i-1$ superscript refers to the value of the variable in the previous time step.

\noindent where $\Delta n_{ser}$ is the serpentine produced at the previous moment, $\Delta m_{ice}$ is the changing mass of ice due to the melting. We calculated the volumes of components from the masses and densities:
\begin{gather}
V = m/\rho \\
V_{voi} = V_{voi}^{i-1}+\frac{3}{10^6}\times\Delta n_{ser}^{i-1} \\
\end{gather}
Amount of substance of water ($n_{wat}$) and olivine ($n_{oli}$):
\begin{gather}
n_{wat} = 1000 \times m_{wat}/ 18 \\
n_{oli} = 1000 \times m_{oli} / 140.71
\end{gather}
%

\subsection{Amount of serpentinite}\label{app:am_serp}
%
To obtain the {\it total} pressure, the sum of the lithospheric ($P_{lit}$) and vapor pressures ($P_{vap}$) we first calculate the gravitational acceleration ($g(l)$) and then $P_{lit}$  in that specific layer:
\begin{gather} \label{eq:P_lit}
\begin{split}
g(l) & = G \frac{m_r}{r^2} \\
P_{lit} & = \sum_{l=r}^{R} \rho_{lay}(l) g(l) dr
\end{split}
\end{gather}
\noindent where $G$ is the gravitational constant, $r$ is the radius of the specific layer, $m_r$ is mass within the radius $r$ , $R$ is the radius of the planetesimal and $dr$ is the thickness of the layer. 

%
From the Clausius-Clapeyron relation, the vapor pressure has the approximate form
\begin{gather}
P_{vap}  = P_0 \, e^{\frac{T_0}{T}}
\end{gather}

\noindent where $P_0=3.58 \times 10^{12}\,Pa$ and $T_0=-6140\,K$ in the presence of ice, while it is changed to $P_0=4.7 \times 10^{10}\,Pa$ and $T_0=-4960\,K$ in the presence of water \citep{Grimm1989}.
In the case of mixed phases of \ce{H2O}, the mass-weighted average of $P_{vap}$ is used \cite{Gobi2017}:
\begin{gather}
\begin{split}
P_{vap}(0) &  = \frac{m_{ice} \, P_0 \, e^{\frac{T_0}{T}} + m_{wat}\,P_0\,e^{\frac{T_0}{T}}}{m_{ice}+m_{wat}} \\
P_{vap} & = \frac{ m_{ice}^{i-1} \, P_0 \, e^{\frac{T_0}{T}} + m_{wat}^{i-1}\,P_0\,e^{\frac{T_0}{T}}}{m_{ice}^{i-1}+m_{wat}^{i-1}}
\end{split}
\end{gather}

The reaction rate depends linearly  on pressure and exponentially on temperature. By knowing the serpentinization rate ($\textit{k}_r$), the number of serpentine produced $\Delta n_{ser}$ can then be calculated in moles:
\begin{gather}
\textit{k}_r = 4383 \,  \frac{P_{lit}+P_{vap}}{10^8} \, e^\frac{-3463}{T} \\
\Delta n_{ser} = min(n_{oli}(1-e^{-kr\times \Delta t}), \frac{n_{wat}}{2})
\end{gather}
\noindent When olivine is in excess, then $n_{wat}/2$ is used as being the limiting reagent.
%
\subsection{Heat budget and temperature increase}\label{app:heat_gain}
To determine the extent of heat production it is necessary to know the heat of reaction:
\begin{gather}
\Delta h_{ser} = \Delta H_r \, \Delta n_{ser}
\end{gather}
\noindent where $\Delta H_r=69\,kJ/mol$ is the reaction enthalpy \cite{Robie1968}

The vaporation heat is calculated as follows:
\begin{gather}
\Delta H_{v} = 3713997.2-7822.6569\,T+17.613373\,T^2-0.019018061\,T^3 \\
\Delta h_{vap} = \Delta H_{v} \, \Delta m_{vap}
\end{gather}
\noindent where $\Delta H_{v}$ is the latent heat of vaporization of water $T$ is capped at 400\,K and $\Delta m_{vap} = m_{vap}-m_{vap}^{i-1}$. 

In those cases when the initial temperature is lower than the melting point of the ice, a certain part of  ice can transform into interfacial water \cite{Anderson1973}.
\begin{gather}
ln\,w = 0.2618+0.5519\,ln\,S-1.449\,S^{-0.264}\,ln(T_{melt}-T)
\end{gather}
\noindent where $w$ is the content of microscopic liquid water (in g/100\,g soil) and $S$ is specific surface area (100\,$m^2\,g^{-1}$ based on \cite{Anderson1973}).

In the next step, we calculated the amount of ice that is converted to water ($\Delta m_{ice}$) considering the actual amount of ice in the layer, the possible amount of interfacial water that can be formed, the amount of water that is able to react in this specific step, and the amount of ice that can be melted by serpentinization.
\begin{gather}\label{eq:meltice}
\begin{split}
\Delta m_{ice}(0) & = min(m_{ice},w \times \frac{m_{oli}+m_{ens}+m_{nre}}{ 100\times \rho_{wat}}) \quad (T\,<\,T_{melt}) \\
            & = \frac{\Delta h_{ser}-\Delta h_{vap}}{L} \quad (T\,>\,T_{melt} \quad and \quad \Delta h_{ser}\,>\,0 ) \\
            & = 0 \quad (T\,>\,T_{melt} \quad and \quad \Delta h_{ser}\,<\,0 ) \\
\Delta m_{ice} & = min(m_{ice},w \times \frac{m_{oli}+m_{ens}+m_{nre}+m_{ser}}{100\,\rho_{wat}}, \\
            & 2\,\Delta n_{ser}\times 0.018), \frac{\Delta h_{ser}-\Delta h_{vap}}{L}) \quad (T\,<\,T_{melt}) \\
			& = \frac{\Delta h_{ser}-\Delta h_{vap}}{L} \quad (T\,>\,T_{melt} \quad and \quad \Delta h_{ser}\,>\,0 ) \\
            & = 0 \quad (T\,>\,T_{melt} \quad and \quad \Delta h_{ser}\,<\,0 )
\end{split}
\end{gather}

\noindent where $L$ is the latent heat of the ice-water phase transition 
\begin{gather}
\Delta h_{ice/wat} = L  \Delta m_{ice}
\end{gather}
The temperature increase ($\Delta T_{ser}$) due to serpentinization is calculated as follows:
\begin{gather}
\begin{split}
\Delta T_{ser}(0) & = \frac{\Delta h_{ser}} {C_p\,(layer)\times m_{lay}} \\
\Delta T_{ser} & = \frac{\Delta h_{ser}-\Delta h_{ice/wat}-\Delta h_{vap}}{C_p\,(layer)\times m_{lay}}
\end{split}
\end{gather}

\subsection{Decay of radionuclides}\label{app:rad}

We considered solely the $^{26}Al$ isotope when calculating the heat from the radiogenic decay.
The radiogenic heat production rate ($Q_{rad}$) is obtained by the following equations \citep{Desch2009}:

\begin{gather}
\begin{split}
\lambda & = \frac{ln(2)}{t_{1/2}} \\
q_{rad} & = TE\times \lambda	\\
Q_{rad} & = q_{rad}\, e^{(-t\times \lambda)} \,m_{rock}\, dt
\end{split}
\end{gather}

\noindent where $\lambda$ is the  decay constant, $TE$ is the total energy and $q_{rad}$ is the radiogenic heat production rate (see Table~\ref{tab:in_val}).

\subsection{Heat transfer}\label{app:Heat trasfer}

\paragraph{Thermal conduction} 
The thermal conductivity of the rocky material components was considered to be independent of the temperature \citep{Cohen2000,Robertson1988}. We calculated the thermal conductivity of the different phases of \ce{H2O} in the following way:
\begin{gather}
\begin{split}
k_{ice} & = 9.828\, exp(-0.0057\,T) \\
k_{vap} & = -0.0143+1.02\times 10^{-4}T \\
k_{wat} & = -0.581 + 6.34 \times 10^{-3}T -7.93\times 10^{-6}T^2 \quad (T\,<\,410\,K) \\
k_{wat} & = 0.9721(-0.142+4.12\times 10^{-3}T-5.01\times 10^{-6}T^2) \quad (T\,>\,410\,K)
\end{split}
\end{gather}

When determining the average thermal conductivity of the investigated layer, a composite rock with a homogeneous material distribution was considered. We calculated the parallel bulk rock conductivity ($k_p$, its grains arranged in a parallel orientation to the direction of heat flow) and the series conductivity ($k_s$, its grains arranged in a layered sequence perpendicular to the heat flow direction). The mean values of $k_p$ and $k_s$ fit well with the observed values, especially for more porous rocks \citep{Robertson1988}:
\begin{gather}
\begin{split}
k_{p} & = nV_1k_1+nV_2k_2+nV_3k_3+... \\
\frac{1}{k_{s}} & = \frac{nV_1}{k_1}+\frac{nV_2}{k_2}+\frac{nV_3}{k_3}+ ... \\
k_{lay} & = \frac{k_p+k_s}{2}
\end{split}
\end{gather}

\noindent where $nV_1$, $nV_2$, $nV_3$ ... are fractional volumes of components ($nV_{component} = V_{component}/V_{lay}$).

\paragraph{Heat radiation} 

On the surface of the body the radiated heat is calculated as:
\begin{gather} \label{eq:sug}
\begin{split}
Q_{th} & = \epsilon \sigma\, (T_{surf}^4-T_{amb}^4) \, A\,dt
\end{split}
\end{gather}
\noindent where $\epsilon$ is the emissivity factor ($\epsilon = 0.9$), $T_{surf}$ the temperature of the surface, $T_{amb}$ is the ambient temperature, which is assumed to be 50\,K, a typical surface temperature of airless bodies due to solar irradiation in the outer Solar System and $A$ is the radiating surface area.

\paragraph{Thermal conduction}

The heat transferred by thermal conduction is calculated as:

\begin{gather} \label{eq:cond}
\begin{split}
Q_{cond} & = A\,k_{lay} \, \frac{\Delta T}{dr}\, dt
\end{split}
\end{gather}
\noindent where $\Delta T$ is the temperature difference between the adjacent layers.

\paragraph{Thermal convection} If the Rayleigh number exceeded a critical value we also calculated the Nusselt number and considered the convective heat flow in our calculations:
\begin{gather}
\begin{split}
Nu & = \left(\frac{Ra}{Ra_{crit}}\right)^{\beta} \\
Q_{conv} & = \frac{k_{lay}}{dr}Nu\,A\,(T(l+1)-T(l))\,dt
\end{split}
\end{gather}
\noindent where $\beta$ is a dimensionless value that can take values between 0.25 and 1/3 depending on the geometry and boundary conditions. As a reference value, we use $\beta$ = 0.3 after \cite{Hussmann2006}. We used Ra$_{crit}$\,=\,1000 as critical Rayleigh number. 

The final heat equation is:
\begin{gather}
Q_{sum}=\Delta h_{ser}-\Delta h_{ice/wat}-\Delta h_{vap}+Q_{rad}-Q_{cool}+Q_{cond}+Q_{conv}
\end{gather}
\noindent where $Q_{cool}=Q_{th}$ in the surface and in the inner layers $Q_{cool}=Q_{cond}$ at the upper boundary of the examined layer.


\end{document}